\documentclass[twocolumn,resetfootnote,tighten]{aastex701}

\usepackage{graphicx}
\usepackage{txfonts} 

\usepackage{natbib}

\usepackage{amsmath,amssymb}

\usepackage{tkz-graph}  
\usetikzlibrary{shapes.geometric, positioning}%

\usepackage{subfig}
\usepackage[shortlabels]{enumitem}

\usepackage{{booktabs}}
\usepackage{rotating}

\newcommand{\totNfields}{91}
\newcommand{\totarea}{$0.3\; \text{deg}^2$}
\newcommand{\totexptime}{500 hours}
\newcommand{\drNfields}{36}

\newcommand{\Msun}{\ifmmode{M_\odot}\else $M_\odot$\xspace\fi}
\newcommand{\MUV}{\ifmmode{M_\textsc{uv}}\else $M_\textsc{uv}$\xspace\fi}
\newcommand{\fesc}{\ifmmode{f_\textrm{esc}}\else $f_\textrm{esc}$\xspace\fi}
\newcommand{\lya}{\ifmmode{\mathrm{Ly}\alpha}\else Ly$\alpha$\xspace\fi}

\begin{document} 
   \title{BEACON: JWST NIRCam Pure-parallel Imaging Survey. III. \\ Constraints on the UV LF and the Clustering of $z\sim7-14$ Galaxies}

    \correspondingauthor{Kimi C. Kreilgaard}
    \email{kimi.cardoso.kreilgaard@nbi.ku.dk}
    
    \author[0009-0005-9953-433X]{Kimi C. Kreilgaard}
    \affiliation{Cosmic Dawn Center (DAWN), Denmark}
    \affiliation{Niels Bohr Institute, University of Copenhagen, Jagtvej 128, DK-2200 Copenhagen N, Denmark}
    \email{kimi.cardoso.kreilgaard@nbi.ku.dk}

    \author[0000-0002-3407-1785]{Charlotte A. Mason}
    \affiliation{Cosmic Dawn Center (DAWN), Denmark}
    \affiliation{Niels Bohr Institute, University of Copenhagen, Jagtvej 128, DK-2200 Copenhagen N, Denmark}
    \email{charlotte.mason@nbi.ku.dk}

    \author[0000-0002-8512-1404]{Takahiro Morishita} 
    \affiliation{IPAC, California Institute of Technology, MC 314-6, 1200 E. California Boulevard, Pasadena, CA 91125, USA}
    \email{takahiro@ipac.caltech.edu}
    \affiliation{Astronomical Institute, Tohoku University, 6-3, Aramaki, Aoba-ku, Sendai, Miyagi 980-8578, Japan}
    
    \author[0000-0003-3817-8739]{Yechi Zhang}
    \affiliation{IPAC, California Institute of Technology, MC 314-6, 1200 E. California Boulevard, Pasadena, CA 91125, USA}
    \email{yechi@ipac.caltech.edu}

    \author[0000-0001-5487-0392]{Viola Gelli}
    \affiliation{Cosmic Dawn Center (DAWN), Denmark}
    \affiliation{Niels Bohr Institute, University of Copenhagen, Jagtvej 128, DK-2200 Copenhagen N, Denmark}
    \email{viola.gelli@nbi.ku.dk}

    \author[0000-0003-4570-3159]{Nicha Leethochawalit}
    \affiliation{National Astronomical Research Institute of Thailand (NARIT), Mae Rim, Chiang Mai, 50180, Thailand}
    \email{nicha@narit.or.th}

    \author[0000-0002-8460-0390]{Tommaso Treu}
    \affiliation{Department of Physics and Astronomy, University of California, Los Angeles, 430 Portola Plaza, Los Angeles, CA 90095, USA}
    \email{tt@astro.ucla.edu}
    
    \author[0000-0001-9391-305X]{Michele Trenti}
    \affiliation{School of Physics, The University of Melbourne, VIC 3010, Australia}
    \email{michele.trenti@unimelb.edu.au}

    \author[0000-0002-5258-8761]{Abdurro'uf}
    \affiliation{Department of Astronomy, Indiana University, 727 East Third Street, Bloomington, IN 47405, USA}
    \email{fnuabdur@iu.edu}

    \author[0000-0002-7570-0824]{Hakim Atek}
    \affiliation{Institut d'Astrophysique de Paris, CNRS, Sorbonne Universit\'e, 98bis Boulevard Arago, 75014, Paris, France}
    \email{atek@iap.fr}

    \author[0000-0001-5984-0395]{Maru\v{s}a Brada{\v c}}
    \affiliation{University of Ljubljana, Department of Mathematics and Physics, Jadranska ulica 19, SI-1000 Ljubljana, Slovenia}
    \affiliation{Department of Physics and Astronomy, University of California Davis, 1 Shields Avenue, Davis, CA 95616, USA}
    \email{marusa.bradac@fmf.uni-lj.si}

    \author[0000-0002-7908-9284]{Larry D. Bradley}
    \affiliation{Space Telescope Science Institute (STScI), 3700 San Martin Drive, Baltimore, MD 21218, USA}
    \email{lbradley@stsci.edu}

    \author[0000-0002-8651-9879]{Andrew J.\ Bunker}
    \affiliation{Department of Physics, University of Oxford, Denys Wilkinson Building, Keble Road, Oxford OX1 3RH, UK}
    \email{Andy.Bunker@physics.ox.ac.uk}

    \author[0009-0009-3404-5673]{Novan Saputra Haryana}
    \affiliation{Astronomical Institute, Tohoku University, 6-3, Aramaki, Aoba-ku, Sendai, Miyagi 980-8578, Japan}
    \email{novan.haryana@astr.tohoku.ac.jp}

    \author[0000-0001-8587-218X]{Matthew J. Hayes}
    \affiliation{Stockholm University, Department of Astronomy and Oskar Klein Centre for Cosmoparticle Physics, AlbaNova University Centre, SE-10691, Stockholm, Sweden}
    \email{matthew.hayes@astro.su.se}

    \author[0009-0002-8965-1303]{Zhaoran Liu}
    \affiliation{Astronomical Institute, Tohoku University, 6-3, Aramaki, Aoba-ku, Sendai, Miyagi 980-8578, Japan}
    \affiliation{MIT Kavli Institute for Astrophysics and Space Research, 70 Vassar Street, Cambridge, MA 02139, USA}
    \email{zhaoran.liu@astr.tohoku.ac.jp}

    \author[0000-0001-7166-6035]{Vihang Mehta}
    \affiliation{IPAC, California Institute of Technology, 1200 E. California Blvd, Pasadena, CA 91125, USA}
    \email{vmehta@ipac.caltech.edu}

    \author[0000-0002-9946-4731]{Marc Rafelski}
    \affiliation{Space Telescope Science Institute, 3700 San Martin Drive, Baltimore, MD, 21218 USA}
    \affiliation{Department of Physics and Astronomy, Johns Hopkins University, Baltimore, MD 21218, USA}
    \email{mrafelski@stsci.edu}

    \author[0000-0002-4140-1367]{Guido Roberts-Borsani}
    \affiliation{Department of Physics \& Astronomy, University College London, London, WC1E 6BT, UK}
    \email{g.robertsborsani@ucl.ac.uk}

    \author[0000-0002-9136-8876]{Claudia Scarlata}
    \affiliation{University of Minnesota, Twin Cities, 116 Church St SE, Minneapolis, MN 55455, USA}
    \email{mscarlat@umn.edu}

    \author[0000-0001-9935-6047]{Massimo Stiavelli}
    \affiliation{Space Telescope Science Institute, 3700 San Martin Drive, Baltimore, MD 21218, USA}
    \email{mstiavel@stsci.edu}

    \author[0009-0005-1487-7772]{Ryo A. Sutanto}
    \affiliation{Astronomical Institute, Tohoku University, 6-3, Aramaki, Aoba-ku, Sendai, Miyagi 980-8578, Japan}
    \email{ryo.sutanto@astr.tohoku.ac.jp}

    \author[0009-0009-8116-0316]{Kosuke Takahashi}
    \affiliation{Astronomical Institute, Tohoku University, 6-3, Aramaki, Aoba-ku, Sendai, Miyagi 980-8578, Japan}
    \email{kosuke.takahashi@astr.tohoku.ac.jp}

    \author[0000-0003-0980-1499]{Benedetta Vulcani}
    \affiliation{INAF -- Osservatorio Astronomico di Padova, Vicolo Osservatorio 5, 35122 Padova, Italy}
    \email{benedetta.vulcani@inaf.it}


\begin{abstract}
The James Webb Space Telescope (JWST) has extended the frontier of galaxy detection to redshifts $z>11$, finding a high abundance of UV-bright sources that challenge theoretical models. However, most current results come from just a few fields, introducing uncertainties due to cosmic variance. 
Here, we constrain $z\sim7-14$ UV luminosity functions (LFs) over $\sim400$ arcmin$^2$ across 36 independent sightlines from DR2 of BEACON, a JWST pure-parallel NIRCam multi-band imaging survey. We identify 164 $7<z<12$ galaxy candidates: 150 F090W-, 14 F115W-, and no robust F150W-dropouts. In the 11 pointings overlapping with public JWST spectroscopy, we find no contaminants, indicating a high purity in our sample.
Our $z\sim7.5$ UV LF agrees with previous bright-end measurements but yields lower number densities at $-21\leq M_\mathrm{UV}\leq-19$. At $z\sim10$, our measurements are lower than most photometric JWST results but match spectroscopic constraints, consistent with the high purity of our selection. The LFs at $z\sim7.5$ and $z\sim10$ are consistent with pre-JWST models, while our limits at $z>13$ do not rule out a possible excess.
We measure significant clustering of bright ($M_\mathrm{UV}<-20.5$) galaxies at $7<z<10$. Fields hosting such sources are approximately three times more likely to be overdense relative to the full survey, implying that UV-bright galaxies preferentially reside in the most massive halos at these redshifts. Comparing with semi-numerical simulations, we estimate that $M_{\mathrm{UV}} < -20.5$ galaxies inhabit halos $\sim0.8$ dex less massive at $z \sim 11$ than at $z \sim 8$, consistent with a shift to higher star formation rates. However, their observed clustering exceeds predictions from pre-JWST luminosity-halo mass relations, suggesting these sources reside in more massive halos than previously modelled and/or multiple halo occupation.
\end{abstract}


\section{Introduction}
In the first few hundred million years after the Big Bang, the first galaxies are predicted to form within the first collapsed dark matter halos \citep[e.g.][]{whiteSimulationsMergingGalaxies1978,Bromm2011}. The \textit{James Webb} Space Telescope (JWST) is now allowing us to observe some of these sources at `Cosmic Dawn'
\citep[see][for a recent review]{starkObservationsFirstGalaxies2026}. One of the most significant early results has been the detection of a higher number density of UV-selected galaxies at $z>10$ than expected based on \textit{Hubble} Space Telescope (HST) and ground-based observations \citep[e.g.][]{castellanoEarlyResultsGLASSJWST2022, bouwensEvolutionUVLF2023, harikaneComprehensiveStudyGalaxies2023, mcleodGalaxyUVLuminosity2023, perez-gonzalezLife30Probing2023, adamsEPOCHSIIUltraviolet2024, donnanJWSTPRIMERNew2024, finkelsteinCompleteCEERSEarly2024,morishitaPhysicalCharacterizationEarly2023, whitler$zGtrsim9$2025}. These observations have proved challenging to explain with pre-JWST theoretical models \citep[e.g.][]{masonGALAXYUVLUMINOSITY2015, tacchellaRedshiftindependentEfficiencyModel2018,yungSemianalyticForecastsJWST2019}, which predict a much steeper decline in the number density of galaxies towards $z\sim14$. This discrepancy between observations and theory implies that either the observations are not representative of the entire Universe or that we are missing ingredients in our galaxy formation models.

These JWST observations have inspired significant theoretical exploration, and various solutions have been proposed to explain the excess of galaxies observed. In general, they pertain to one of two classes: (i) Either galaxies in the early Universe are, on average, brighter than expected, which could be a consequence of more efficient star formation \citep{inayoshiLowerBoundStar2022, dekelEfficientFormationMassive2023, liFeedbackfreeStarburstsCosmic2024, somervilleDensitymodulatedStarFormation2025}, a top-heavy initial mass function (IMF) \citep{cuetoASTRAEUSIXImpact2024, trincaExploringNatureUVbright2024, hutterASTRAEUSIndicationsTopheavy2025}, or less dust obscuration \citep[e.g.][]{ferraraStunningAbundanceSuperearly2023,fioreDustywindclearJWSTSuperearly2023}, or (ii) galaxies in the early Universe exhibit more bursty star-formation, so that galaxies in more numerous low mass dark matter halos can be temporarily several magnitudes brighter than average \citep{masonBrightestGalaxiesCosmic2023,Sun2023,shenImpactUVVariability2023, gelliImpactMassdependentStochasticity2024}.
However, it remains unclear which of these mechanisms dominates.
Furthermore, most early JWST results on the UV LF came from just a handful of legacy fields, meaning they suffer from systematic uncertainties due to significant cosmic variance \citep{Trenti2008,Trapp2020,asadaEarliestGalaxyEvolution2026}.

Tracing the evolution of the galaxy population in the first billion years thus requires imaging searches over a wide area, ideally in many independent fields to minimise cosmic variance. HST already began to achieve this kind of science through pure-parallel surveys such as BoRG \citep{trentiBRIGHTESTREIONIZINGGALAXIES2011}, HIPPIES \citep{yanPROBINGVERYBRIGHT2011} and WISP \citep{atekWFC3INFRAREDSPECTROSCOPIC2010}. Now, thanks to the sensitivity of JWST, pure-parallel NIRCam imaging surveys are providing an efficient way to achieve wide-area high-redshift candidate searches across numerous independent pointings. In Cycle 1, the PANORAMIC survey provided NIRCam imaging spanning 1.1-5.0\,$\mu$m \citep{williamsPANORAMICSurveyPure2025}, enabling the selection of $z>10$ candidates in 28 independent fields. Combining PANORAMIC data with legacy fields, \citet{weibelExploringCosmicDawn2026} demonstrated that the $z\sim10-14$ galaxy excess persists even in these fields where the cosmic variance is minimised. In Cycle 2, the BEACON pure-parallel survey \citep{Morishita2023_beacon} obtained 0.8-5.0\,$\mu$m NIRCam imaging, also enabling $z\sim7-9$ candidate selection -- providing the opportunity to self-consistently track the evolution of the UV LF in the redshift range where JWST and HST selections of Lyman-break galaxies overlap. These pure-parallel surveys are also proving ideal for finding rare bright galaxy candidates \citep{donnanSpectroscopicConfirmationLarge2026}.

However, even with robust estimates of the UV LF it is still challenging to distinguish theoretical models, as, by construction, both types of proposed scenarios to explain the excess in number density of early galaxies (described above) produce similar galaxy UV LFs at $z\sim10-14$.
A potential key to breaking these degeneracies is to estimate the underlying dark matter halo masses of the galaxies we are observing \citep{Ren2018,Mirocha2020b}.
The observed clustering of galaxies has long been used to link galaxies to their dark matter halos \citep[see e.g.][for a review]{Wechsler2018}.
As recently discussed by \citet{munozBreakingDegeneraciesFirst2023} and \citet{gelliImpactMassdependentStochasticity2024}, a shift to more bursty star formation at high redshift would decrease the expected clustering of UV bright galaxies, as bright galaxies are more likely to be hosted in low mass halos.
Clustering measurements of photometrically-selected galaxies at $z\sim4-7$ from Subaru and HST imaging have demonstrated that bright galaxies ($\MUV \simeq -22$) tend to be more clustered than faint ($\MUV \simeq -19$) galaxies \citep{Harikane2016,qiuDependenceGalaxyClustering2018,harikaneComprehensiveStudyGalaxies2023,Barone-Nugent2014a}, implying they reside in more massive halos.

JWST has extended clustering estimates to $z>7$ for the first time, providing a path to break degeneracies between galaxy evolution models. 
Initial results from measuring angular correlation functions have suggested an increase in galaxy bias at fixed selection with increasing redshift at $z\lesssim 11$ \citep{dalmassoGalaxyClusteringCosmic2024, dalmassoAcceleratedEvolutionGalaxy2026, paquereauTracingGalaxyhaloConnection2025}, suggesting a continued close correlation between galaxy luminosity (or stellar mass) and halo mass, potentially favouring the increased luminosity scenarios above.
However, angular correlation function measurements with JWST remain challenging due to the small area of deep legacy fields, and potential contamination in high-redshift selections in wider area surveys such as COSMOS-Web \citep{shuntovCOSMOSWebStellarMass2025}, limiting our ability to constrain the halo masses of the brightest $z>7$ galaxies.
Pure-parallel surveys have long been recognised as a powerful alternative to probe galaxy clustering as they provide wide areas to search for UV-bright galaxies and randomly sample the matter density field  \citep{Robertson2010,trentiBRIGHTESTREIONIZINGGALAXIES2011,cameronObservationalDeterminationGalaxy2019}. 
\citet{weibelExploringCosmicDawn2025a} recently used count-in-cells statistics \citep{Adelberger1998,Robertson2010} to estimate galaxy bias at $z \sim 10$ in the PANORAMIC Cycle 1 JWST pure-parallel survey, finding evidence for large cosmic variance. 
While this demonstrated the potential of pure-parallels for clustering estimates, the analysis was restricted to a single redshift epoch due the NIRCam filter configuration. Furthermore, this work demonstrated that the small NIRCam field-of-view limits our ability isolate the linear clustering signal when using count-in-cells statistics.
Progress requires extending clustering estimates from pure parallel observations to a broader redshift range to quantify how the galaxy-halo connection evolves over the crucial $z\sim7-12$ range where the LF appears to differ from pre-JWST predictions, and exploring alternative statistics which can account for non-linear statistics in pure-parallel observations.

In this paper, we present an analysis of $z>7$ galaxy candidates from DR2 of BEACON: the JWST Cycle 2 NIRCam pure-parallel survey \citep[GO-3990, PIs: Morishita, Mason, Trenti, Treu;][]{morishitaBEACONJWSTNIRCam2025}\footnote{\url{https://beacon-jwst.github.io/}}, spanning 91 NIRCam pointings. We use \drNfields\ BEACON fields ($\sim400$\,arcmin$^2$) with sufficient NIRCam coverage for robust selection of $z\sim7-18$ Lyman-break galaxies, providing an updated estimate of the UV LF with minimal cosmic variance, including the widest area survey for F090W-dropouts ($z\sim7.5$, 306\,arcmin$^2$) with JWST to-date. We leverage the unbiased nature of BEACON and the sensitivity of NIRCam imaging to investigate the environments of UV-bright galaxies at $z>7$. Through comparison with simulations, we link the excess overdensity probability of fields hosting UV-bright galaxies to their typical halo mass.

The paper is structured as follows. In Sect.~\ref{sec:data}, we present the BEACON survey, while the high-redshift candidate selection is described in Sect.~\ref{sec:selection}. Section~\ref{sec:uvlfs} details the fitting of the UV luminosity functions at $z>7$, including a description of the completeness simulations and measurement of the number densities in three redshift bins: $z\sim7.5$, $z\sim10$ and $z>13$. Section~\ref{sec:overdensity} characterises the density of the \drNfields\, independent BEACON DR2 fields and assesses the environment of fields hosting the brightest galaxies. We discuss our results in Sect.~\ref{sec:disc} and present our conclusions in Sect.~\ref{sec:conclusion}.

We adopt the AB magnitude system \citep{okeSecondaryStandardStars1983}, and cosmological parameters: $\Omega_m=0.3$, $\Omega_\Lambda=0.7$, and $H_0 = 70$ $\mathrm{km s}^{-1}\mathrm{ Mpc}^{-1}$.

\section{Data}
\label{sec:data}
This study is based on high-redshift photometric galaxy candidates from the JWST Cycle 2 pure-parallel NIRCam imaging survey, BEACON \citep[PID: 3990,][]{morishitaBEACONJWSTNIRCam2025}. The survey is described in Sect.~\ref{sub:survey_overview}, while Sect.~\ref{sub:survey_configs} provides an overview of the filter and exposure configurations chosen for the observations.

\subsection{Survey Overview}
\label{sub:survey_overview}
BEACON is a large JWST Cycle 2 program that provides pure-parallel multi-band NIRCam imaging of \totNfields\;independent sightlines, totalling approximately \totarea\;and \totexptime\;of on-source exposure time.

This work uses \drNfields\;BEACON pointings, forming the second data release (DR2)\footnote{The data set is available on the survey website: \url{https://beacon-jwst.github.io}}, which presents multi-band imaging with a minimum of six broadband filters suitable for $z>7$ galaxy selections. The first data release (DR1), using the first 19 fields, was presented by \citet{morishitaBEACONJWSTNIRCam2025}. We refer the reader to that paper for more details on the survey design and the data reduction strategy. Here, we briefly summarise the survey design and present various statistics from the BEACON DR2 data set.

One main goal of the BEACON survey was to mitigate the effects of cosmic variance thanks to its random pointing (pure-parallel) nature and significant total area covered. The JWST pure-parallel observing mode allows the use of an instrument (in our case NIRCam) while another instrument is used as the primary observing mode. Since the primary programs determine the pointing, we cannot obtain completely random pointings. However, since most targets are in the foreground relative to our $z>7$ search in this study, BEACON offers a sampling that is close to unbiased. Of course, there are exceptions, owing to popular areas of the sky that are known to contain overdensities (e.g. Abell-2744 that was recently confirmed to host an overdensity at $z=10$ by \citealp{castellanoEarlyResultsGLASSJWST2023} and \citealp{napolitanoSevenWondersCosmic2025}). Still, pure-parallel observations provide an efficient way to reduce the effects of cosmic variance \citep[][]{Trapp2020,williamsPANORAMICSurveyPure2025,morishitaBEACONJWSTNIRCam2025,roberts-borsaniBoRGJWSTSurveyProgram2025}.

\begin{figure*}
    \centering
    \includegraphics[width=\linewidth]{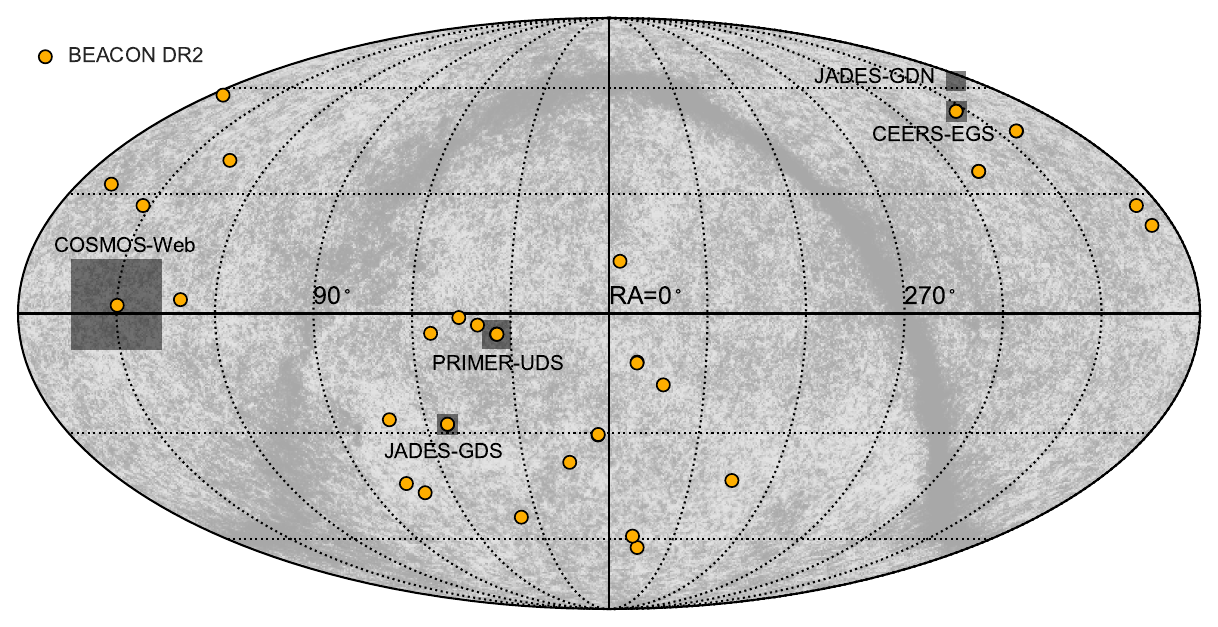}
    \caption{Sky distribution of the \drNfields\, fields included in BEACON DR2 (orange circles), overlaid on a temperature map from the WMAP 7 yr data \citep{jarosikSEVENYEARWILKINSONMICROWAVE2011}. We also show some JWST legacy fields (grey squares): COSMOS-Web \citep{caseyCOSMOSWebOverviewJWST2023}, PRIMER-UDS \citep{dunlopPRIMERPublicRelease2021}, CEERS-EGS \citep{finkelsteinCosmicEvolutionEarly2025}, JADES-GOODS-South, and JADES-GOODS-North \citep{eisensteinOverviewJWSTAdvanced2026}. The symbol sizes do not represent the true sky size, but they are roughly scaled to the corresponding survey area.}
    \label{fig:skymap}
\end{figure*}

In Fig.~\ref{fig:skymap} we show the distribution of the \drNfields\;BEACON DR2 fields on the sky, each covering one NIRCam field of view of 9.7 arcmin$^2$, resulting in a total effective area of  392\,arcmin$^2$ for the full survey so far. For comparison, we also show the position and observed area for some legacy fields covered by JWST (COSMOS, UDS, EGS,
GOODS-South, and GOODS-North). BEACON DR2 provides 25 completely new fields, which have not been observed by JWST NIRCam before. 11 BEACON DR2 fields overlap with fields with existing JWST observations, enhancing the legacy value of these datasets and enabling validation of candidate selection in our NIRCam-only fields, e.g. via additional photometric filter coverage and spectroscopy. 

\subsection{Filter and Exposure Configuration}
\label{sub:survey_configs}
Due to its parallel nature, the BEACON survey is characterised by inhomogeneous exposure and filter configurations across the different fields \citep[see][for extensive discussion]{williamsPANORAMICSurveyPure2025,morishitaBEACONJWSTNIRCam2025}. Full descriptions of BEACON's filter configuration are given in \citet{morishitaBEACONJWSTNIRCam2025} and \citet{zhangBEACONJWSTNIRCam2026}, while we provide a summary of details relevant to our study here.

The standard filter configuration for BEACON consists of 8 NIRCam filters spanning the wavelength range $0.8~\mu\text{m} < \lambda < 5.0~\mu\text{m}$ (F090W, F115W, F150W, F200W, F277W, F356W, F410M, F444W). Due to the unpredictable availability of parallel observing `slots' for prime observations in Cycle 2, in some cases, fewer filters were used. Conversely, when possible, more filters (usually medium bands) were used. If the available parallel slots for one opportunity offered different exposure times, we allocated the longest exposure to slots consisting of the filter pairs (F090W, F410M) and (F115W, F444W) to secure a non-detection blueward of the Lyman break in $z>7$ galaxies and a secure detection of the continuum in F444W.

In this study, we use all fields observed with at least 6 NIRCam filters in BEACON, typically:
F090W, F115W, F150W, F277W, F356W, and F444W, enabling dropout selection of Lyman-break galaxies at $z\sim7-14$, although for a handful of fields F200W was configured over F090W or F150W (see Tab.~\ref{tab:field_depths}).
Our sample consists of a total of \drNfields\, fields with at least 6 NIRCam filters, 11 fields with at least 8 NIRCam filters and 5 fields with 10 or more filters.

\begin{figure}
    \centering
    \includegraphics[width=\columnwidth]{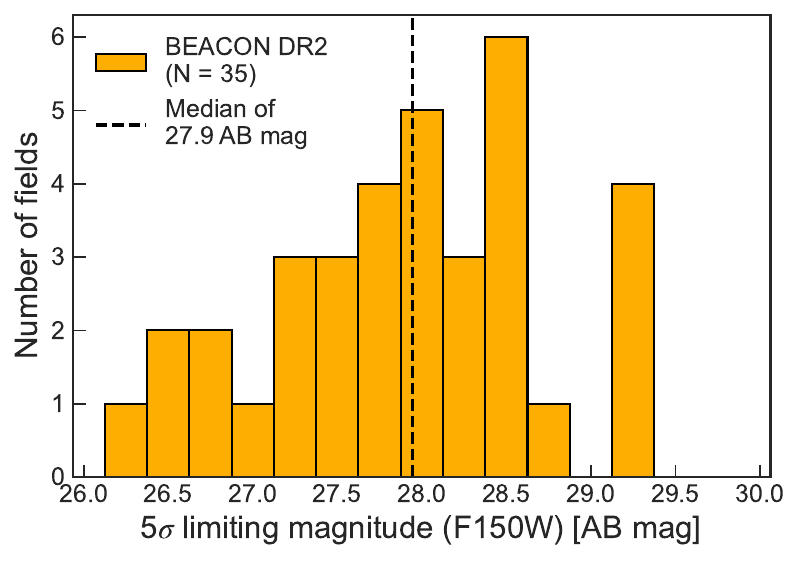}
    \caption{Distribution of the $5\sigma$ limiting magnitude (depth) measured within a $0.\!''16$ aperture radius in the F150W band, present for 35 of the \drNfields\, fields included in this study (BEACON DR2). The median depth of 27.9 AB mag across the 35 fields is marked with a dashed line.}
    \label{fig:depths}
\end{figure}

Table~\ref{tab:field_depths} provides an overview of the 5$\sigma$ depths measured within a $0.\!''16$ aperture radius (in each filter) for all \drNfields\ fields, which form BEACON DR2 and are used in this study.
Figure~\ref{fig:depths} shows the distribution of 5$\sigma$ depths in F150W across these fields. We find a median depth of 27.9 mag, but note that several fields reach $\sim$1.5 mag deeper.

BEACON DR1 \citep{morishitaBEACONJWSTNIRCam2025}, consisting of the first 19 pointings, is contained within DR2. Thus, in Tab.~\ref{tab:field_depths}, we also note which fields double as DR1 fields. Future work in DR3 will combine all BEACON observations, including an additional 55 pointings with fewer than 6 filters, and archival data.

\section{High Redshift Candidate Selection}
\label{sec:selection}
In this work, we focus on measuring the $z>7$ galaxy UV LF; thus, we first need to identify a sample of high-redshift galaxy candidates from BEACON DR2. This work presents the number densities and overdensity characterisation of this $z>7$ sample, while the physical properties of these galaxies are described by \citet{zhangBEACONJWSTNIRCam2026}

The selection presented here is extracted from imaging data reduced using the same procedure as \citet{morishitaBEACONJWSTNIRCam2025}. In Sect.~\ref{sub:photometry}, we describe how we identified sources and estimated their photometric redshift. Section~\ref{sub:dropout-selection} details how the final sample of candidates was selected, including a description of the dropout selections employed and the subsequent visual inspection.

\subsection{Photometry and Photometric Redshifts}
\label{sub:photometry}
For each of the \drNfields\, fields in BEACON DR2, we identified potential sources and performed multi-band photometry using \texttt{Source Extractor} \citep[SE;][]{bertinSExtractorSoftwareSource1996}.
We follow the same procedure as used in BEACON DR1 \citep{morishitaBEACONJWSTNIRCam2025}, and thus refer to this DR1 paper and the accompanying DR2 paper, \citet{zhangBEACONJWSTNIRCam2026}, for more details on the imaging reduction, source detection and photometry recovery. Below, we briefly outline the method.

The imaging data reduction is performed on the raw images retrieved from the Mikulski Archive for Space Telescopes (MAST) and reduced with the official JWST pipeline, with the addition of the following operations. We perform $1/f$-noise subtraction with \texttt{bbpn}\footnote{\url{https://github.com/mtakahiro/bbpn}} \citep{takahiromorishitaMtakahiroBbpnV132023}, snowball-masking with \texttt{Grizli} \citep{brammerGrizli2022}, and cosmic-ray masking with \texttt{lacosmic} \citep{vandokkumCosmicRayRejectionLaplacian2001,larrybradleyLarrybradleyLacosmic1102023}. The final drizzled images, resampled to $0.\!''0315$\,/\,pixel, are aligned to the IR-detection image, i.e. a variance-weighted coadd of the F277W, F356W, and F444W filters. This filter combination ensures robust detection of both high-redshift galaxies and red, dust-obscured or quiescent galaxies. Before measuring fluxes, all NIRCam mosaics are PSF-matched to the F444W resolution. We generate convolution kernels using PSFs produced with 
\texttt{stpsf}\footnote{\url{https://github.com/spacetelescope/stpsf}}, and pass those to \texttt{pypher} \citep{boucaudConvolutionKernelsMultiwavelength2016} to produce convolution kernels for each filter.

We then run SE in dual-image mode, to identify sources in the IR-detection image and measure fluxes in each PSF-matched filter. The SE parameters match those used in \citet{morishitaBEACONJWSTNIRCam2025} and \citet{zhangBEACONJWSTNIRCam2026}. 
Aperture fluxes are measured in fixed circular apertures of radius $0.\!''16$ to optimise S/N while minimising PSF-dependent systematics. Following \citet{morishitaBEACONJWSTNIRCam2025}, we convert these aperture fluxes to total fluxes by applying a single multiplicative factor per source, defined as the ratio $f_{\rm auto,F444W}/f_{\rm aper,F444W}$, where $f_{\rm auto}$ is the total flux measured with a scaling factor of 2.5, estimated to enclose $\sim94\%$ of the total flux \citep{bertinSExtractorSoftwareSource1996}.
Photometric uncertainties are corrected for pixel–pixel correlations introduced during the drizzling process: we determine an empirical noise-rescaling factor by measuring fluxes in 300 randomly placed empty apertures and matching their standard deviation to the median SE-reported errors. Finally, all fluxes are corrected for galactic extinction using the line-of-sight reddening values from NED \citep{schlegelMapsDustInfrared1998,schlaflyMEASURINGREDDENINGSLOAN2011} and adopting the \citet{cardelliRelationshipInfraredOptical1989} Milky Way extinction curve. More details of the imaging reduction, PSF homogenisation, and photometric calibration are presented in \citet{morishitaBEACONJWSTNIRCam2025} and \citet{zhangBEACONJWSTNIRCam2026}.

We estimate the photometric redshift of the extracted sources above by running the SED-fitting code \texttt{EAZY} \citep{brammerEAZYFastPublic2008} with the updated template library presented in \citet{hainlineCosmosItsInfancy2024}, similar to the procedure described in \citet{morishitaBEACONJWSTNIRCam2025} employed for BEACON DR1. However, here we also supplement the template library with the one presented in \citet{naiduSchrodingersGalaxyCandidate2022}, which captures the spectral signature of a galaxy with dusty SFE at low redshift that can otherwise mimic a high-redshift Lyman-break galaxy. Together, these additional templates supplement the standard \texttt{EAZY} library (v1.3) so we can better capture the signatures of young, line-emitting galaxies and exclude low-redshift contaminants, thus enabling better redshift determination in the early Universe. The redshift posteriors, $p(z)$, obtained from this fitting procedure are utilised in the dropout selection described in Sect.~\ref{sub:dropout-selection} below.

\subsection{Dropout Selection and Final Sample}
\label{sub:dropout-selection}

We adopt the selection approach of BEACON DR1 \citep{morishitaBEACONJWSTNIRCam2025} \citep[see also][]{morishitaBrightendGalaxyCandidates2018,Roberts-Borsani2022b,morishitaEnhancedSubkiloparsecscaleStar2024}. This is a two-step selection process, whereby we first select sources based on a Lyman-break dropout, then apply a cut on the photometric redshift distribution to minimise contamination from low-redshift interlopers.

We select dropouts in F090W, F115W and F150W, which allow us to identify candidates at $7 \lesssim z \lesssim 18$. 
The F090W selection covers a similar redshift range to F850LP dropouts with HST/WFC3 and ACS \citep[see e.g.][]{bunkerContributionHighredshiftGalaxies2010,oesch7GALAXIESHUDF2010,wilkinsUltravioletPropertiesStarforming2011,lorenzoniStarformingGalaxies2011} and the F115W selection covers a similar redshift range to F105W and F125W dropouts from HST, enabling a comparison between HST and JWST number densities.
Our selection does not explicitly include a colour criterion,
allowing us to preserve sources which may have UV colours outside traditional selections due to features like strong UV emission lines, or reddening due to e.g. nebular continuum, dust or high HI column densities which have been demonstrated in JWST spectra \citep[e.g.][]{cameronNebularDominatedGalaxies2024, roberts-borsaniExtremesJWSTSpectroscopic2024, roberts-borsaniJWSTSpectroscopicInsights2026, tangJWSTSpectroscopicProperties2026, asadaImprovingPhotometricRedshifts2025}. 

For each source, we require at least a $2\sigma$ non-detection in the dropout filter and all available filters blueward. Additionally, we require at least a $4\sigma$-detection in the first filter redward of the Lyman break (excluding the filter covering the break), i.e. where the source is expected to be brightest. 
For example, for F115W-dropouts, which should exhibit a Lyman break within F150W, we require that the source be observed with F150W, but there are no requirements on the flux in this filter.
Instead, we require $<2\sigma$ detections in all available filters blueward of the F150W filter (i.e. F090W, F115W for the standard filter configuration, but also in F070W if available) and a $>4\sigma$ detection in the first filter redward of F150W (typically in F200W, but for the few fields where F200W is not available, F277W is used instead).

To improve the purity of our sample, we use two additional requirements on the \texttt{EAZY} photometric redshift distributions to remove low-redshift interlopers: (i) For each of the three dropout selections, we only include candidates with a high probability of being at a high redshift. We thus require that $p(z_\mathrm{phot}>z_{\rm min}) > 80\%$ for a candidate to remain in the selection, where $z_{\rm min}$ for each dropout selection is defined below. (ii) We require that the high-redshift solution is strongly preferred by imposing a maximum difference in the $\chi^2$ between the best high- and low-redshift fits, such that: 
$\chi^2_{\rm{high-z}} - \chi^2_{\rm{low-z}} < -4$.

Finally, to avoid overlap between the different dropout selections, we impose an upper limit on the best-fit photometric redshift, $z_{\rm best}$, derived from \texttt{EAZY}. Specifically, we require $z_{\rm best} < z_{\rm max}$, where $z_{\rm max} = 9.7$, $13.0$, and $18.0$ for the F090W-, F115W-, and F150W-dropout selections, respectively, ensuring that candidates are uniquely assigned to a selection.

Below, we state the signal-to-noise (S/N) requirements on the relevant filters and $z_{\rm min}$ in each selection for our standard filter configuration.

\begin{itemize}
    \item[$\blacksquare$] \textbf{F090W-dropouts ($7.2 \lesssim z < 9.7$)} \\
        $$S/N_{\rm 150} > 4 $$
        $$S/N_{\rm 090} < 2$$
        $$z_{\rm min} = 6$$
    \item[$\blacksquare$] \textbf{F115W-dropouts ($9.7 \lesssim z < 13.0$)} \\
        $$S/N_{\rm 200} > 4 $$
        $$S/N_{\rm 115, 090} < 2$$
        $$z_{\rm min} = 8$$
    \item[$\blacksquare$] \textbf{F150W-dropouts ($13.0 \lesssim z < 18$)} \\
        $$S/N_{\rm 277} > 4 $$
        $$S/N_{\rm 150, 115, 090} < 2$$
        $$z_{\rm min} = 10$$
\end{itemize}

\begin{deluxetable*}{ccccc}
\tablecaption{Final galaxy candidates of BEACON DR2 \label{tab:candidates}}
\tablewidth{0pt}
\tablehead{
\colhead{Dropout filter} &
\colhead{Redshift window} &
\colhead{Number of eligible fields} &
\colhead{Effective survey area} &
\colhead{Number of final candidates}
}

\startdata
F090W & $7.3 \lesssim z \lesssim 9.7$ & 30 & 306.4 arcmin$^2$ & 150 \\
F115W & $9.7 \lesssim z \lesssim 13$ & 35 & 370.5 arcmin$^2$ & 14 \\
F150W & $13 \lesssim z \lesssim 18$ & 19 & 216.0 arcmin$^2$ & $0^\dagger$ \\
\enddata

\tablenotetext{}{The number of final galaxy candidates remaining in each of the three dropout selections after applying our selection criteria and visually inspecting the sources. Shown is also the number of BEACON DR2 fields with the necessary filter coverage to meet the requirements of the selection, and the effective survey area of those fields excluding bad pixels. $^\dagger$One F150W-dropout candidate was initially identified but excluded from the final sample due to its non-detection in independent overlapping datasets, suggesting it is unlikely to be a genuine high-redshift galaxy (more details in Sect.~\ref{sub:dropout-selection}).}

\end{deluxetable*}

This selection resulted in a total of 266 candidates. Each candidate was then visually inspected by three of the co-authors (KK, YZ, TM) to exclude data artefacts with poor photometry resulting from extended diffraction spikes from stars, detector edges, bad pixels, and/or hot pixels originating from cosmic rays. After visual inspection, we are left with a total of 165 galaxy candidates at $z>7$, consisting of 150 F090W-dropouts, 14 F115W-dropouts, and one F150W-dropout source.

In Tab.~\ref{tab:candidates}, we show for each of the three dropout samples: the number of fields included that meet the filter configuration requirements such that the given selection can be performed, the effective survey area where the selection is attempted, and the number of final galaxy candidates, i.e. sources that meet the requirements described above and pass the visual inspection.

\begin{figure}
    \centering
    \includegraphics[width=\linewidth]{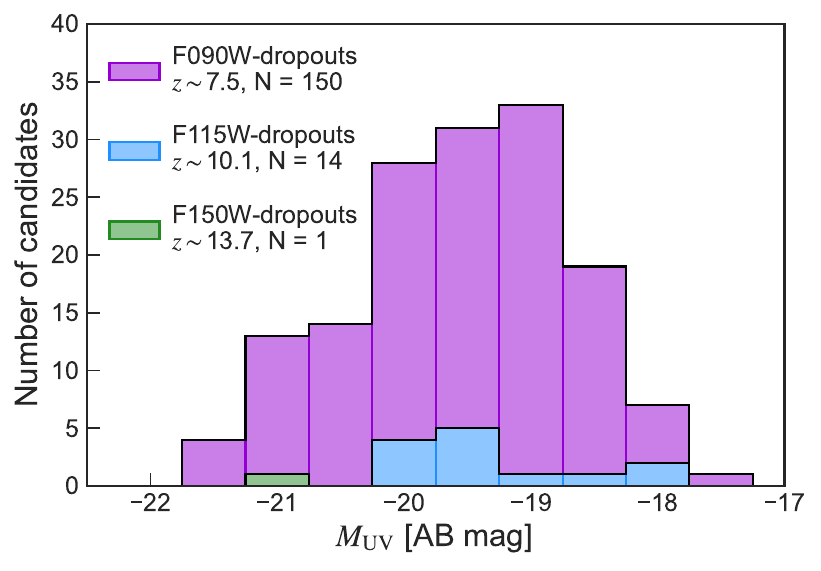}
    \caption{Histogram showing the absolute UV magnitudes for galaxy candidates in each of the three dropout selections. The UV magnitudes are measured as the average flux in the rest-frame 1450\AA--1550\AA\ wavelength range based on a power law fitted to the UV continuum sampled by the available NIRCam photometry for each source. We note that the F150W-dropout candidate is not included in our final sample, as its non-detection in independent overlapping datasets suggests it is likely not a galaxy (see Sect.~\ref{sub:dropout-selection}).}
    \label{fig:Muv_hist}
\end{figure}

In Fig.~\ref{fig:Muv_hist}, we display a histogram of the absolute UV magnitudes, $M_\mathrm{UV}$, recovered for the final sample of galaxies. We estimate $M_\mathrm{UV}$ as the average flux in the rest-frame 1450\AA--1550\AA\ wavelength range based on a power law fitted to the UV continuum sampled by the available NIRCam photometry for each source. The fit is performed in wavelength space assuming $f_\lambda \propto \lambda^\beta$, using filters selected according to each source’s \texttt{EAZY} photometric redshift (see Sect.~\ref{sub:photometry}) to probe the rest-frame UV continuum while avoiding potential Ly$\alpha$ contamination. The reported $M_\mathrm{UV}$ values and errors (see Tab.~\ref{tab:f090w_dropouts_1}--\ref{tab:f115w_dropouts_1}) are computed with Monte Carlo simulations, sampling 200 realisations per source of the NIRCam photometry and the \texttt{EAZY} photometric redshift to properly account for the propagated uncertainties. We refer to \citet{zhangBEACONJWSTNIRCam2026} for more details on measuring the absolute UV magnitudes of the BEACON DR2 candidates.

\begin{figure}
    \centering
    \includegraphics[width=\linewidth]{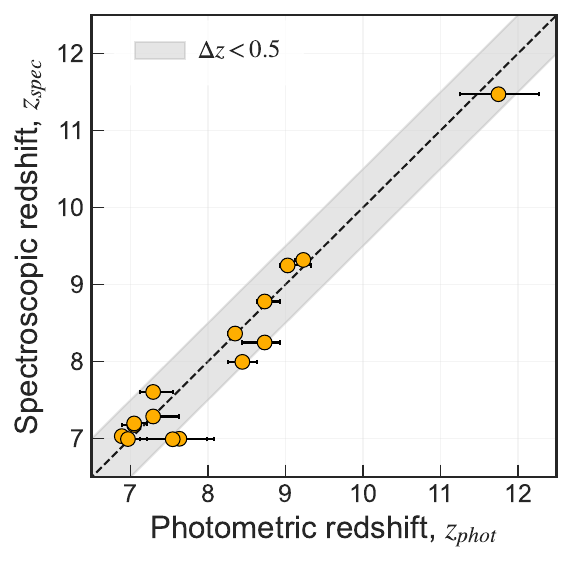}
    \caption{Comparison between the photometric redshift, as estimated from the \texttt{EAZY} SED fit and the spectroscopic redshift when available. The orange points represent the median of the \texttt{EAZY} redshift posterior, while the lower and upper errors indicate the 16th and 84th percentiles. The dashed line indicates the one-to-one relation, while the shaded region shows the $\Delta z < 0.5$ margin.}
    \label{fig:zspec}
\end{figure}

To assess the purity of our sample, we compared our photometric redshifts to spectroscopic redshifts in BEACON fields overlapping with public JWST spectroscopy. We recover spectroscopic redshifts by matching sources based on position with the \texttt{v4.4} spectroscopic catalogue from the DAWN JWST Archive (DJA)\footnote{\url{https://dawn-cph.github.io/dja/}}, requiring \texttt{grade=3} to ensure reliable spectroscopic measurements.  
In the 11 BEACON DR2 pointings with overlapping public NIRSpec coverage (in Abell-2744, UDS, GOODS-S, and EGS), we find zero low-redshift contaminants and 15 of our candidates are spectroscopically confirmed with excellent agreement between photometric and spectroscopic redshifts, as shown in Fig.~\ref{fig:zspec}. 
We list the spectroscopic confirmations and their origin in the candidate list in Tables~\ref{tab:f090w_dropouts_1}--\ref{tab:f115w_dropouts_1}.

This corresponds to an observed purity of 100\% based on the final candidates in our sample that have spectroscopic coverage. Treating these 15 matches as independent Bernoulli trials, this translates to a lower limit on the purity rate of $>82\%$ at a 95\% confidence level. The spectroscopic matches span an absolute magnitude range of approximately $-22 \leq M_\mathrm{UV} \leq -19$ and redshifts in the range $7<z<11.5$, which encapsulates the properties of the vast majority of galaxy candidates in our final sample and thus indicates that the high purity is representative of our full sample. The matches are found in fields with 6--16 filters and thus also portray the inhomogeneous filter coverage present in BEACON fields. However, while the BEACON fields with spectroscopic matches present varying depths, they are mostly found in fields with median depth or deeper imaging, as these are more likely to overlap with legacy fields with spectroscopic coverage. For example, the F150W $5\sigma$ limiting magnitudes range between 28.0 and 29.2 AB mag in the 6 fields containing spectroscopic matches,  i.e. deeper than the median of 27.9 AB mag across the 36 fields in BEACON DR2 (see Fig.~\ref{fig:depths}). Nonetheless, the 17 DR2 fields with F150W-imaging deeper than 28 mag account for 131/150 and 12/14 of the recovered sources in respectively the F090W- and F115W-dropout selection, corresponding to 87\% of our final sample.

Our estimated lower limit of $>82\%$ on the purity rate based on the spectroscopic sample is relatively conservative compared to the $5\text{--}10\%$ interloper fraction reported by \citet{leethochawalitUVRestframeBand2026} for their $z\sim8$ photometric sample. In that work, the number of low-redshift contaminants was estimated for nine public JWST extragalactic fields observed in Cycle 1 \citep{morishitaEnhancedSubkiloparsecscaleStar2024}, with F150W $5\sigma$ depths ranging from 28.0 to 29.8 mag, comparable to or up to $\sim$0.5 mag deeper than our data. Their analysis used the JAGUAR templates of low-redshift galaxies \citep{williamsJWSTExtragalacticMock2018}, adding field-specific photometric noise to mock galaxies and applying selection criteria broadly similar to those adopted here.

To assess the completeness of our sample, we search the DJA catalogue for all spectroscopically confirmed $z\geq7$ sources with imaging in BEACON DR2, where our selections can be attempted. We find 28 spectroscopically confirmed galaxies, including the 15 successfully recovered sources, as described above.
Of the remaining 13 galaxies, 7 do not pass our S/N selection criteria as they have insufficient S/N in detection bands due to shallower BEACON imaging compared to earlier imaging of these sources, while the other 6 do not pass our $p(z)$ requirements. Of these 6, 3 sources appear robust in BEACON photometry but are just below our $p(z)$ thresholds, and the remaining 3 have catastrophic SEDs that are not well-fit by \texttt{EAZY}, including one potential little red dot (LRD) \citep{mattheeLittleRedDots2024a,kocevskiRiseFaintRed2025}. These missed sources will be accounted for in our completeness simulations; see Sect.~\ref{sub:completeness} below.

We also note, as described by \citet{morishitaBEACONJWSTNIRCam2025}, our selection successfully selects 12/13 of the spectroscopically confirmed galaxies at $z>10$ known at the end of 2024 \citep[the only unsuccessful case is a very faint galaxy, $m_{\rm F200W} \approx 29$,][for which we recover a very broad $p(z)$ and thus would not pass our $p(z)$ cut]{Curtis-Lake2022}.

\begin{figure*}
    \centering
    \includegraphics[width=\textwidth]{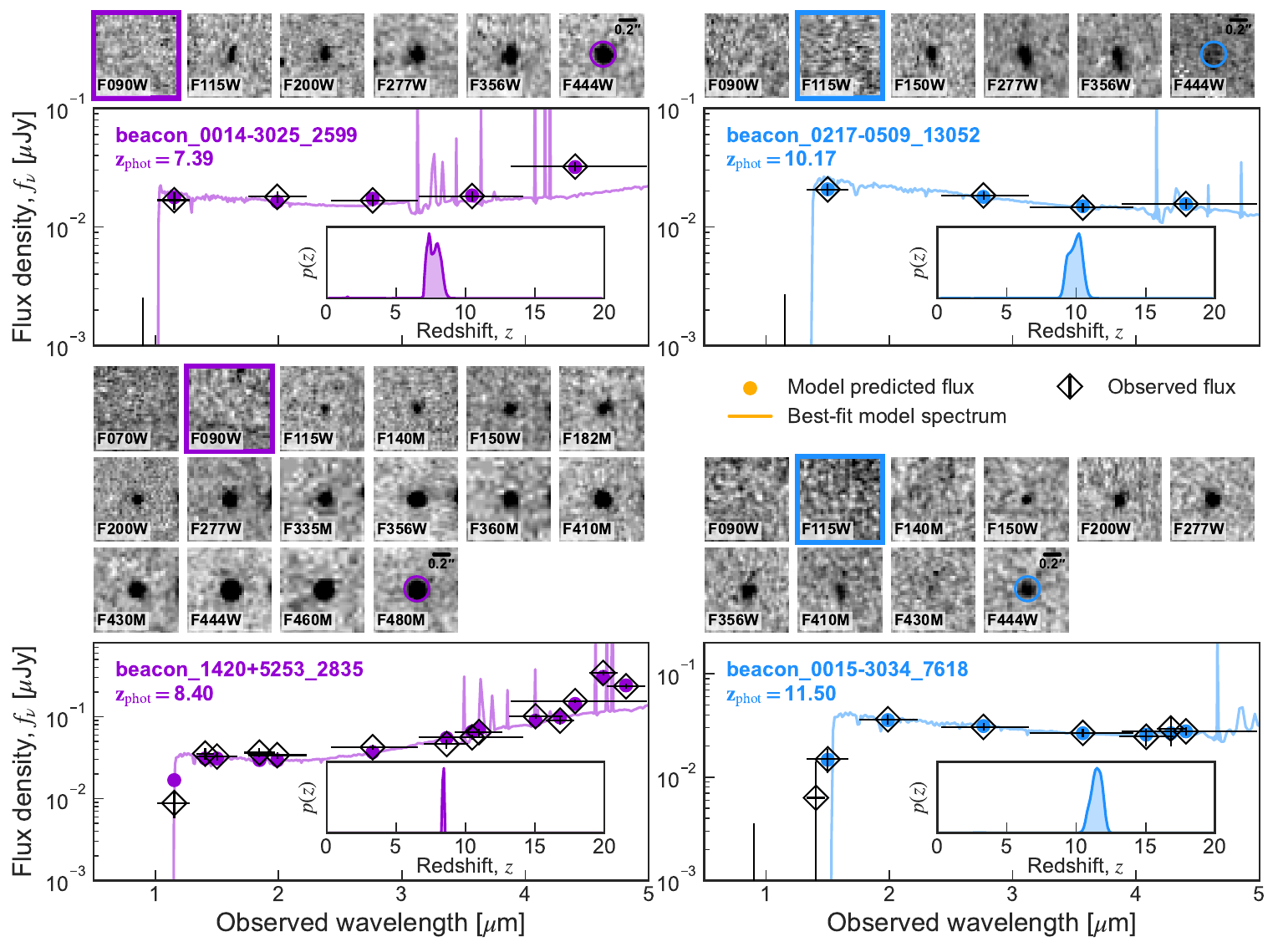}
    \caption{Examples of our observed photometry and SED fitting results for two F090W-dropouts (purple) and two F115W-dropouts (blue). Black open diamonds show the observed photometry in each band with errors, where the uncertainty range on the wavelength axis is determined from the half-power wavelengths of the passband, e.g. the wavelengths at which the transmission falls to 50\% of its peak value. The coloured curves are the best-fit SEDs from \texttt{EAZY}, while the coloured circles indicate the corresponding predicted photometry. The insert panel shows the marginalised redshift posterior distribution, $p(z)$, from the SED fitting. We also show image cutouts for each available filter for the given source, marking the dropout filter with a coloured border. The $0.\!''16$ circular aperture in which fluxes are measured is shown on the F444W cutout.}
    \label{fig:SEDs}
\end{figure*}

Finally, in Fig.~\ref{fig:SEDs}, we show examples of the recovered sources, displaying the best-fit \texttt{EAZY} SED model, along with the observed flux in each band and the predicted flux from the model SED.

We recover a $z_\mathrm{phot}=13.69^{+ -0.03}_{-0.32}$ F150W-dropout source, \texttt{beacon\_1420+5253\_4770} \citep[see][for a discussion of its physical properties]{zhangBEACONJWSTNIRCam2026}. This source was also independently recovered by  \citet{weibelExploringCosmicDawn2026} (their ID 89475) with a similar photometric redshift. With a UV magnitude of $M_\mathrm{UV}=-21.19^{+0.08}_{-0.09} \text{\,mag}$, it is one of the brightest candidates detected at $z>13$ to date. This source was found in the \texttt{beacon\_1420+5253} field, which is located within the EGS legacy field. There is a strong break located between the F150W and F182M band, where we detect the source with S/N=7.8, thus indicating a high photometric redshift \citep{zhangBEACONJWSTNIRCam2026}. However, as noted by \citet{donnanSpectroscopicConfirmationLarge2026}, this source was not detected in overlapping CEERS (GO-1345, PI: S. Finkelstein; \citealp[]{finkelsteinCosmicEvolutionEarly2025}). We also verify that it is not detected in SAPPHIRES imaging (GO-6434, PI: E. Egami; \citealp[]{sunSlitlessArealPureParallel2025}). 
The transient nature of this source will be discussed by Toshikage et al. (in prep), and we exclude it from the sample used in this study.
A full estimate of the expected rate of these transients and its impact on purity of high-redshift selections is beyond the scope of this work, but will be an important consideration for future pure-parallel surveys \citep[see also,][]{DeCoursey2025}.

\section{UV Luminosity Functions}
\label{sec:uvlfs}
Here, we present the UV luminosity function for each of the three dropout selections. In Sect.~\ref{sub:completeness} we describe our completeness simulations. Section~\ref{sub:LF_densities} details how the effective volumes and the number densities are measured, and Sect.~\ref{sub:LF_fitting} describes the UV LF fitting procedure. In Sect.~\ref{sub:LF_results}, we report our findings for the number densities and the best-fit UV LFs. Finally, in Sect.~\ref{sub:LF_models}, we compare our fitted UV LFs to models and compute the luminosity density.

\subsection{Completeness}
\label{sub:completeness}

Due to the inhomogeneous filter and exposure configurations characterising the pure-parallel BEACON fields, we perform individual completeness simulations for each field. 

We employ the adaptation of the code \texttt{GLACiAR2} \citep{leethochawalitUVLuminosityFunctions2023,leethochawalitQuantitativeAssessmentCompleteness2021}, presented by \citet{morishitaBEACONJWSTNIRCam2025}, to determine the completeness and selection function for each field. This method injects galaxies with a Sérsic profile of $n = 1$, following the $M_\mathrm{UV}$-size relation of \citet{morishitaEnhancedSubkiloparsecscaleStar2024} and template SEDs from the JAGUAR catalogues \citep{williamsJWSTExtragalacticMock2018}. We refer the reader to \citet{morishitaBEACONJWSTNIRCam2025} for more details, and here only describe differences between our simulations and those in DR1.

\begin{figure*}
    \centering
    \includegraphics[width=\textwidth]{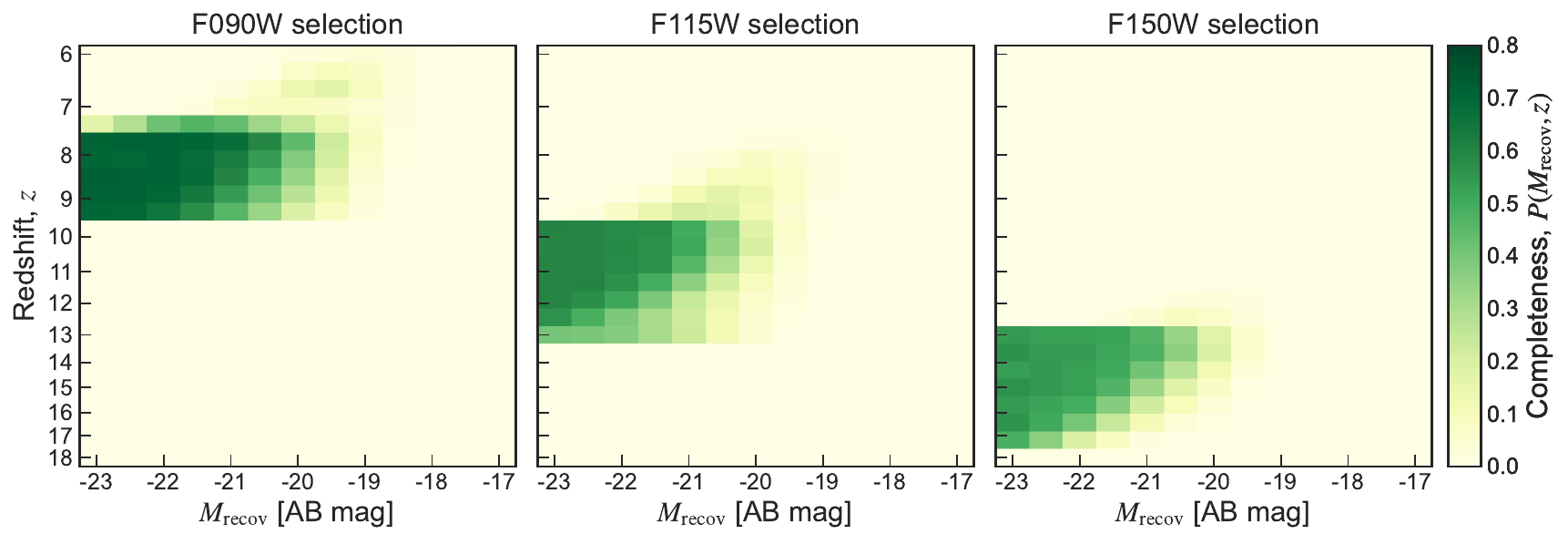}
    \caption{Example of the completeness matrix for each of the three dropout selections, as a function of the recovered magnitude, $M_\mathrm{recov}$, and the injected/intrinsic redshift, $z$. Shown are results of the completeness simulations for the field \texttt{beacon\_2304-6250}, which is a typical BEACON DR2 field, with a F150W 5$\sigma$ limiting magnitude of 27.94 AB mag (i.e. roughly the median of 27.9 mag for DR2). This field has been observed with 8 filters: F090W, F115W, F150W, F200W, F277W, F356W, F410M, and F444W.}
    \label{fig:completeness}
\end{figure*}

The method and choice of parameters are identical to those described by \citet{morishitaBEACONJWSTNIRCam2025}, except for the following. In each field, for each intrinsic magnitude and redshift bin, we inject a fixed density of galaxies, such that one galaxy is injected per 30 square arcseconds in each field, excluding zero or bad pixels. This corresponds to injecting $1100\text{--}3000$ galaxies per field. The UV magnitude bins range between $M_\mathrm{UV}=-16.5\;\text{mag}$ and $-23.0\;\text{mag}$ with an increment of 0.5. The redshift bins range between $z=6$ and 18, with a total of 24 bins that get progressively wider towards higher redshift, where objects are rarer.

We perform source detection, photometric extraction, and selection of the injected galaxies similar to how we select real sources (as described in Sect.~\ref{sec:selection}). Here, however, we use the F444W band as the detection image due to the functional limitation of the software. Since the injected galaxies are not affected by galactic extinction, we do not apply an extinction correction to their fluxes, ensuring a self-consistent treatment.

We can now define the completeness function (for each field) in one redshift bin as a function of the recovered and intrinsic magnitude of the injected galaxies \citep{leethochawalitQuantitativeAssessmentCompleteness2021}:
\begin{equation}
    P(M_\mathrm{recov}, M_\mathrm{in}, z) = \frac{N(M_\mathrm{recov}, M_\mathrm{in}, z)}{N(M_\mathrm{in}, z)}.
    \label{eq:full_completeness}
\end{equation}
Here, $N(M_\mathrm{recov}, M_\mathrm{in}, z)$, denotes the number of galaxies in a redshift bin, $z$, with recovered UV magnitude, $M_{\rm{recov}}$, and intrinsic UV magnitude, $M_{\rm{in}}$, while $N(M_\mathrm{in}, z)$ counts all galaxies injected with the UV magnitude $M_{\rm{in}}$, regardless of their recovered magnitude. Notice, our completeness function, $P(M_\mathrm{recov}, M_\mathrm{in}, z)$, combines both the detection completeness and selection function into one. 

In Fig.~\ref{fig:completeness} we show an example of the recovered selection window and completeness (as a function of the recovered magnitude and the redshift, see Eq. \ref{eq:our_completeness}) for a typical field in our sample, the field \texttt{beacon\_2305-6250}, which has 8 filter coverage and reaches a F150W 5$\sigma$ depth of 27.94 mag, roughly the median for BEACON DR2 fields (Fig.~\ref{fig:depths}). At the brightest UV magnitudes and central redshift region of the dropout selections, we reach, respectively, 72\%, 60\%, and 56\% completeness for the F090W-, F115W-, and F150W-dropout selections for this example field. Failure to recover injected galaxies occurs either when a source is not detected, for example, due to blending with existing sources, or when it fails to satisfy our selection criteria, typically because of the noise artificially added to the fluxes.

\subsection{Measuring the Number Density of Galaxies}
\label{sub:LF_densities}

We estimate the number densities of galaxies in fixed magnitude bins for each of the three dropout selections of galaxy candidates (see Sect.~\ref{sub:dropout-selection}), following e.g. \citet[][]{morishitaBrightendGalaxyCandidates2018, morishitaBEACONJWSTNIRCam2025}.

We define the completeness (see also Eq.~\ref{eq:full_completeness}) for each dropout selection as a function of only the recovered absolute UV magnitudes \citep[Method 2 in][]{leethochawalitQuantitativeAssessmentCompleteness2021}:
\begin{equation}
   P(M_\mathrm{recov}, z) = 
   \frac{\sum_{M_\mathrm{in}}
   N(M_\mathrm{recov}, M_\mathrm{in}, z)}
   {N(M_\mathrm{in} \in M_\mathrm{recov}, z)}.
    \label{eq:our_completeness} 
\end{equation}
This represents the ratio of the number of galaxies recovered in one $M_\mathrm{UV}$ bin to the number of galaxies with intrinsic magnitude in the same $M_\mathrm{UV}$ bin, for a given redshift. Here, the numerator sums all galaxies with the indicated recovered magnitude, regardless of their intrinsic magnitude. Depending on the overlap between the sample of injected and recovered magnitudes, this definition of the completeness function can thus, in principle, exceed unity.

The effective comoving volume of one field of view (FoV) thus also becomes a function of the recovered absolute UV magnitude \citep[see e.g.][]{steidelLymanBreakGalaxies$zgtrsim$1999,leethochawalitQuantitativeAssessmentCompleteness2021}:
\begin{equation}
    V_\mathrm{eff}(M_\mathrm{recov}) = \int_{0}^{\infty} \frac{dV}{dz} P(M_\mathrm{recov}, z) \; dz.
    \label{eq:eff_vol}
\end{equation}
Here, $dV/dz$ is the comoving, differential volume, using the area of the observed field, excluding bad pixels.
We compute the integral numerically over the redshift range probed by the completeness simulation ($z=6\text{--}18$), and match the $M_\mathrm{UV}$ bins we measure number densities in by linearly interpolating the completeness for each redshift bin, $P(M_\mathrm{recov})$, to the desired magnitude grid.

Finally, the number density, $n_\mathrm{obs}$, in a given $M_\mathrm{UV}$ bin with width $\Delta M_\mathrm{UV}$ is computed as:
\begin{equation}
    n_\mathrm{obs}(M_\mathrm{recov}) = 
    \frac{ N_\mathrm{obs}(M_\mathrm{recov}) }
    {\sum_{N_\mathrm{fields}} V_\mathrm{eff,i}(M_\mathrm{recov}) \cdot \Delta M_\mathrm{UV}},
    \label{eq:gal_density}
\end{equation}
where $N_\mathrm{obs}(M_\mathrm{recov})$ is the number of observed galaxies across all fields that pertain to the given selection and have an observed absolute UV magnitude within the $M_\mathrm{recov}$ bin. We divide by the total effective volume: a sum over the effective volumes of the individual fields. We only include fields where the given selection was attempted (See Tab.~\ref{tab:candidates}) in this sum to ensure we only account for the probed volume in the calculation of the number densities.

One field (\texttt{beacon\_0014-3025}) is located near a massive foreground cluster, Abell-2744, and we therefore correct for the magnification using the lens model by \citet{bergaminiGLASSJWSTEarlyRelease2023}. Most sources are not strongly lensed, exhibiting magnification in the range $\mu \sim 1 \text{--} 2$. Thus, for objects in this field, we de-magnify their absolute UV magnitudes: $M_\mathrm{int} = M_\mathrm{obs} + 2.5 \cdot \log_{10}(\mu)$. Finally, we divide the effective area of this field by the median magnification of sources in the field, $\bar{\mu}=1.77$ to recover the intrinsic effective area. In blank fields, gravitational lensing (magnification bias) is not expected to significantly impact the UV LF at $M_\mathrm{UV} \gtrsim -22$ \citep{masonCORRECTING8GALAXY2015}, so we do not correct other sources for magnification.

In Tab.~\ref{tab:number_densities}, we list the effective volume, the number of objects, and the resulting number densities in each bin, for each dropout selection. Errors on the number densities are reported as the one-sided 84\% confidence interval of a Poisson distribution for upper limits (corresponding to the 1$\sigma$ limit for a Gaussian distribution) when no objects were detected, and the two-sided limit otherwise \citep[]{gehrelsConfidenceLimitsSmall1986}. 
We further discuss these number densities and compare with other works in Sect.~\ref{sub:LF_results}.

\begin{deluxetable}{cccc}
\tablecaption{Number densities of galaxies at $z>7$\label{tab:number_densities}}

\tablehead{
\colhead{$M_{\mathrm{UV}}$} &
\colhead{Effective volume} &
\colhead{Number of} &
\colhead{Number density} \\
\colhead{[mag]} &
\colhead{$[10^{3}\,\mathrm{Mpc}^3]$} &
\colhead{objects} &
\colhead{$[\log_{10}(\phi\,\mathrm{Mpc}^{-3}\,\mathrm{mag}^{-1})]$}
}
\startdata
\multicolumn{4}{c}{F090W-dropouts ($z_\mathrm{median}=$7.47)} \\
\hline
$-23$ & $898.05$ & $0$ & $<-5.69$ \\
$-22$ & $902.40$ & $2$ & $-5.65^{+0.45}_{-0.63}$ \\
$-21$ & $826.24$ & $17$ & $-4.69^{+0.15}_{-0.17}$ \\
$-20$ & $562.63$ & $53$ & $-4.03^{+0.09}_{-0.09}$ \\
$-19$ & $184.09$ & $62$ & $-3.47^{+0.08}_{-0.08}$ \\
$-18$ & $12.92$ & $16$ & $-2.91^{+0.16}_{-0.17}$ \\
$-17$ & $0.27$ & $0$ & $<-2.17$ \\
\hline
\multicolumn{4}{c}{F115W-dropouts ($z_\mathrm{median}=$10.14)} \\ \hline
$-23$ & $1206.75$ & $0$ & $<-5.82$ \\
$-22$ & $1080.06$ & $0$ & $<-5.77$ \\
$-21$ & $795.39$ & $1$ & $-5.90^{+0.62}_{-1.08}$ \\
$-20$ & $385.04$ & $8$ & $-4.68^{+0.23}_{-0.26}$ \\
$-19$ & $84.61$ & $2$ & $-4.63^{+0.45}_{-0.63}$ \\
$-18$ & $4.35$ & $3$ & $-3.16^{+0.37}_{-0.48}$ \\
$-17$ & $0.40$ & $0$ & $<-2.34$ \\
\hline
\multicolumn{4}{c}{F150W-dropouts 
($z > 13$)} \\ \hline
$-23$ & $620.69$ & $0$ & $<-5.53$ \\
$-22$ & $510.07$ & $0$ & $<-5.44$ \\
$-21$ & $319.27$ & $0$ & $<-5.24$ \\
$-20$ & $130.50$ & $0$ & $<-4.85$ \\
$-19$ & $24.62$ & $0$ & $<-4.13$ \\
$-18$ & $2.60$ & $0$ & $<-3.15$ \\
$-17$ & $0.73$ & $0$ & $<-2.60$ \\
\enddata

\tablenotetext{}{The binned number densities measured for the three selections: F090W- ($z\sim7.5$), F115W- ($z\sim10$), and F150W-dropouts ($z\sim14$). For the magnitude bins with detections, we report the $1\sigma$ confidence interval, while for bins with no detections, we report the $1\sigma$ upper limit. We also display the effective volume computed based on the completeness simulations and the number of objects in each magnitude bin.}

\end{deluxetable}

\subsection{Fitting the UV Luminosity Function}
\label{sub:LF_fitting}
Here, we fit UV luminosity functions to the number counts of objects in fixed magnitude bins for the two dropout selections where we have detections, i.e. F090W- and F115W-dropouts. We test two parameterisations of the UV LF commonly used at high redshifts: a Schechter function \citep[see e.g.][]{bouwensEvolutionUVLF2023} and a Double power law (DPL) \citep[see e.g.][]{Bowler2020,donnanJWSTPRIMERNew2024}, and fit them following common procedures in the literature \citep[e.g.][]{leethochawalitUVLuminosityFunctions2023,whitler$zGtrsim9$2025}.

With the \citet{schechterAnalyticExpressionLuminosity1976} parameterisation, the UV LF as a function of the intrinsic absolute magnitude, $\phi(M_\mathrm{in})$, is given by:
\begin{equation}
    \begin{aligned}
     \phi(M_\mathrm{in} \mid\phi^*, M^*, \alpha) = &\frac{\phi^*\ln(10)}{2.5} \cdot 10^{-0.4\cdot(M_\mathrm{in}-M^*)\cdot(\alpha+1)} \\
     &\cdot \exp\left(-10^{-0.4\cdot(M_\mathrm{in}-M^*)} \right),
     \end{aligned}
     \label{schechter}
\end{equation}
while for the double power law, it is:
\begin{equation}
    \begin{aligned}
    \phi(M_\mathrm{in} \mid\phi^*, M^*, \alpha, \beta) = 
    &\frac{\phi^*}
    {10^{0.4(M_\mathrm{in}-M^*)(\alpha+1)} +
    10^{0.4(M_\mathrm{in}-M^*)(\beta+1)}}
    \end{aligned}
    \label{DPL}
\end{equation}
Here $\phi^*$ is the normalisation, $M^*$ is the characteristic magnitude, $\alpha$ is the faint-end slope, and $\beta$ is the bright-end slope (present only in the DPL function).

To fit the parameterisations above, we use the affine-invariant, ensemble, Markov chain Monte Carlo (MCMC) sampler \texttt{emcee} \citep{foreman-mackeyEmceeMCMCHammer2013} with the default \texttt{StretchMove}. We use 100 walkers, each performing 10,000 sampling steps, with initial positions uniformly sampled from the respective parameter priors. We discard the first 1000 steps of each chain as burn-in steps and extract every 10th step as posterior samples to ensure independent samples of the free parameters.

For the F090W-dropout selection, we assign uniform priors on all parameters within the following ranges: 
$\log_{10} ( \phi^* \cdot \mathrm{Mpc}^{3} \mathrm{mag}) \in [-8,1]$, where $\phi^*$ is in units $~\text{Mpc}^{-3}\mathrm{mag}$, 
$M^* \in [-24;-18]~\text{mag}$, 
$\alpha \in [-3;-1]$,
and $\beta \in [-6;-2]$ (for the DPL). Towards higher redshifts, there are fewer detections, especially at faint $M_\mathrm{UV}$, and thus less constraining power. For the F115W-dropout selection, we therefore limit the range on the $\alpha$-prior to $\alpha \in [-3;-2.1]$, such that the upper limit on the faint-end slope corresponds roughly to the 84th percentile of the marginalised posterior in the F090W-dropout fit for $\alpha$, keeping the other priors the same, following similar approaches in the literature. 

We adopt a Poissonian log-likelihood to evaluate the probability of detecting $N_\mathrm{data}$ galaxies with a recovered absolute UV magnitude within a certain $M_\mathrm{UV}$ bin, given the sampled model parameters:
\begin{equation}
    \ln(\mathcal{L}) = \sum_{M_\mathrm{recov}} \left[ N_\mathrm{data}\cdot \ln(N_\mathrm{model}) - N_\mathrm{model} - \ln(N_\mathrm{data}!) \right],
    \label{eq:log-likelihood}
\end{equation}
where $N_\mathrm{model}$ denotes the corresponding expected number as predicted from the sampled UV LF parameters, accounting for the completeness. We calculate $N_\mathrm{model}$ via:
\begin{equation}
    N_\mathrm{model}(M_\mathrm{recov}) = 
    \phi(M_\mathrm{recov}) \cdot \Delta M_\mathrm{UV}
    \sum_{N_\mathrm{fields}} V_\mathrm{eff,i}(M_\mathrm{recov}),
    \label{eq:model_equation}
\end{equation}
where $\phi(M_\mathrm{recov})$ is the sampled UV LF evaluated at $M_\mathrm{recov}$, $\Delta M_\mathrm{UV}$ is the bin width that the observed number counts have been measured in, and $V_\mathrm{eff,i}(M_\mathrm{recov})$ is the effective volume of each field, estimated using Eq.~\ref{eq:eff_vol}, for the given magnitude bin.

The inferred UV LF model parameters for the F090W- and the F115W-dropout selections and the two parameterisations are summarised in Tab.~\ref{tab:uvlf_params}. In Sect.~\ref{sub:LF_results}, we comment on the inferred UV LF fits and compare with other observations.

\begin{deluxetable*}{ccccccc} 
\tablecaption{Fitted UVLF parameters and luminosity density
\label{tab:uvlf_params}} 
\tabletypesize{\small} 
\renewcommand{\arraystretch}{1.2} 
\tablehead{ 
\colhead{Redshift bin} & 
\colhead{Model} & 
\colhead{$\phi^*$} & 
\colhead{$M^*_\mathrm{UV}$} & 
\colhead{$\alpha$} & 
\colhead{$\beta$} & 
\colhead{$\rho_\mathrm{UV}(M_\mathrm{UV}<-17)$} \\ 
\colhead{} & 
\colhead{} & \colhead{$(10^{-5}\,\mathrm{Mpc}^{-3}\,\mathrm{mag}^{-1})$} & \colhead{(mag)} & 
\colhead{} & 
\colhead{} & \colhead{$(10^{25}\,\mathrm{erg}\,\mathrm{s}^{-1}\,\mathrm{Hz}^{-1}\,\mathrm{Mpc}^{-3})$} }

\startdata
F090W dropouts & Schechter & $3.55$$^{+5.97}_{-2.75}$ & $-21.27$$^{+0.55}_{-0.82}$ & $-2.18$$^{+0.10}_{-0.26}$ & -- & $2.48$$^{+0.42}_{-0.25}$ \\ 
$z\sim7.5$ & DPL & $1.90^{+23.36}_{-1.54}$ & $-21.39^{+1.54}_{-1.09}$ & $-2.31$$^{+0.25}_{-0.26}$ & $-4.54^{+1.89}_{-0.27}$ & $2.57$$^{+0.46}_{-0.28}$ \\ 
\hline
F115W dropouts & Schechter & $2.25^{+0.55}_{-2.22}$ & $-20.35^{+0.28}_{-2.11}$ & $-2.16^{+-0.09}_{-0.62}$ & -- & $0.45^{+0.31}_{-0.07}$ \\ 
$z\sim10$ & DPL & $0.65^{+2.09}_{-0.64}$ & $-20.77^{+0.86}_{-2.05}$ & $-2.41^{+0.06}_{-0.49}$ & $-5.91^{+3.23}_{--0.76}$ & $0.47^{+0.29}_{-0.11}$ \\ 
\hline
\hline
\enddata

\tablenotetext{}{The fitted parameters for the two parameterisations of the UV LF (Schechter and DPL functions) for the two dropout selections where we have detections. We report the best-fit parameter value, while errors are reported as the 68\% highest density posterior (HDP) interval. We note that for some parameters, the fit is not well-constrained, and the best-fit value falls outside of the HDP region. We also report the corresponding UV luminosity density obtained by integrating down to a faint-end limit of $M_\mathrm{UV}<-17$, where the errors are estimated from the 16th and 84th percentiles of the luminosity densities computed from the UV LF resulting from 1000 realisations of the MCMC parameter samples.}
\end{deluxetable*}

\subsection{UV LF Results and Comparison with Other Observations}
\label{sub:LF_results}

\begin{figure*}
    \centering
    \includegraphics[width=\textwidth]{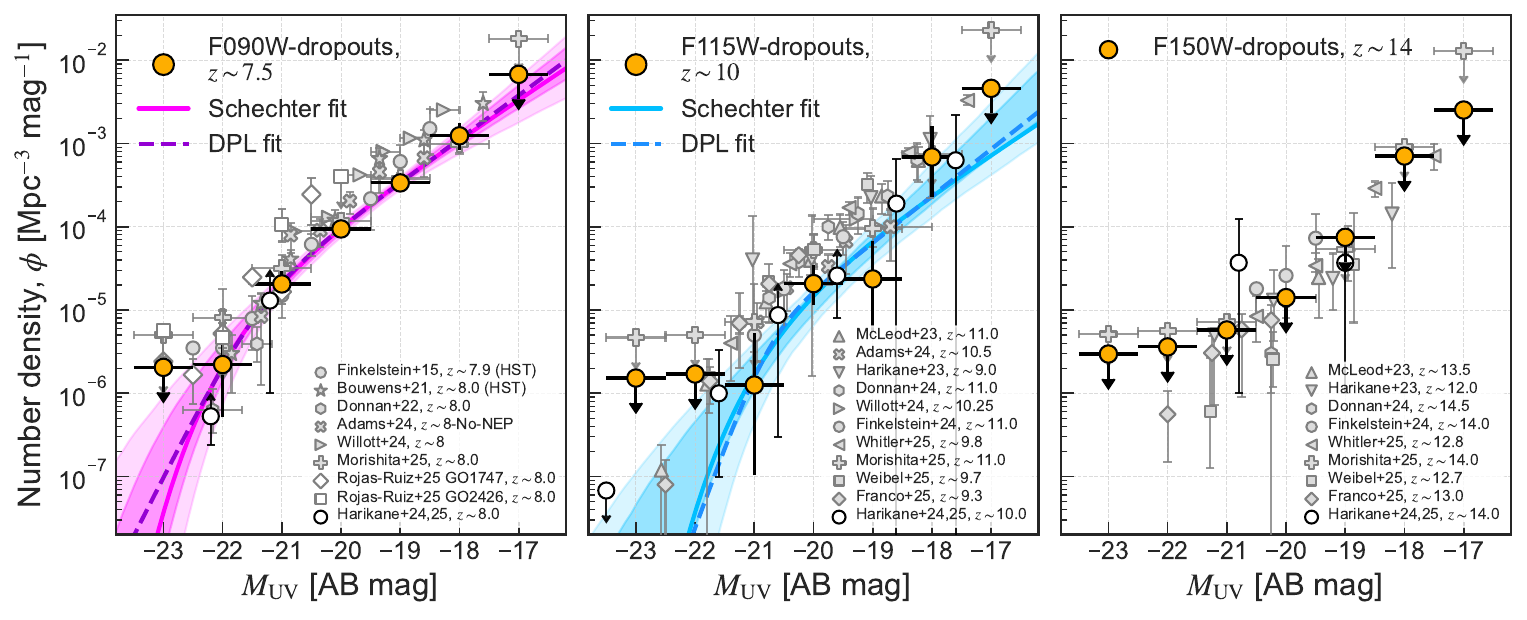}
    \caption{Binned UV LFs inferred from our F090W- (left panel), F115W- (middle panel), and F150W-dropout samples (right panel),
    The orange scatter points indicate the number density of galaxies in fixed $M_\mathrm{UV}$ bins. The best-fit Schechter- and DPL parameterisations of the UV LF are displayed in, respectively, the solid and dashed coloured lines, while the shaded regions indicate the 68\% and 95\% confidence intervals of the Schechter fits (the confidence intervals of the DPL fits are similar to the Schechter ones, though a little higher at the bright end). For comparison, we also plot number densities based on photometry by \citet[]{finkelsteinEVOLUTIONGALAXYRESTFRAME2015, bouwensNewDeterminationsUV2021, donnanEvolutionGalaxyUV2022, mcleodGalaxyUVLuminosity2023, harikaneComprehensiveStudyGalaxies2023,willottSteepDeclineGalaxy2024, finkelsteinCompleteCEERSEarly2024, donnanJWSTPRIMERNew2024, adamsEPOCHSIIUltraviolet2024, morishitaBEACONJWSTNIRCam2025, whitler$zGtrsim9$2025, weibelExploringCosmicDawn2026} and \citet[]{francoPhysicalPropertiesGalaxies2025} (grey symbols), all of which probe survey areas of $\sim50~\text{arcmin}^2$ or larger. In addition, we plot number densities based on a purely spectroscopic sample by \citet{harikanePureSpectroscopicConstraints2024,harikaneJWSTALMAKeck2025} and based on a spectroscopically-informed photometric sample by \citet{rojas-ruizBoRGJWSTSurveyAbundance2025} (white scatter points).}
    \label{fig:uvlfs}
\end{figure*}

In Fig.~\ref{fig:uvlfs}, we plot the number densities measured in Sect.~\ref{sub:LF_densities} and reported in Tab.~\ref{tab:number_densities}, together with the best-fit UV luminosity functions inferred in Sect.~\ref{sub:LF_fitting} and reported in Tab.~\ref{tab:uvlf_params}.
The error bars on our number densities represent the two-sided 84\% confidence limit of a Poisson distribution, while the upper limits are estimated as one-sided limits (both corresponding to 1$\sigma$ limits for a Gaussian distribution). The solid (dashed) lines are the Schechter (DPL) UV LF curves defined by the set of sampled parameters with the highest log-likelihood in the MCMC sampling, for each dropout selection. The shaded regions indicate the 68\% (and 95\%) confidence intervals for the Schechter parametrisation, estimated as the grid-wise 16th and 84th (2.5th and 97.5th) quantiles of the UV LF curves resulting from 1000 realisations of the MCMC parameter samples. The confidence intervals for the DPL fits are similar. 
We find no clear preference for either the Schechter function or the double power law parametrisation of the LFs. However, since we detect no galaxies brighter than $M_\mathrm{UV}<-22$ in any of the redshift bins, the bright-end slope, $\beta$, used in the DPL parametrisation is not well constrained.

Overall, we find a decline in number densities with increasing redshift ($z\sim7-14$), broadly consistent with trends in the literature. 
For comparison, we also plot number densities, based on photometry, reported in the literature from HST by \citet{finkelsteinEVOLUTIONGALAXYRESTFRAME2015} and \citet{bouwensNewDeterminationsUV2021}, and from JWST by \citet[]{donnanEvolutionGalaxyUV2022, mcleodGalaxyUVLuminosity2023, harikaneComprehensiveStudyGalaxies2023,willottSteepDeclineGalaxy2024, finkelsteinCompleteCEERSEarly2024, donnanJWSTPRIMERNew2024, adamsEPOCHSIIUltraviolet2024, morishitaBEACONJWSTNIRCam2025, whitler$zGtrsim9$2025, weibelExploringCosmicDawn2026} and \citet[]{francoPhysicalPropertiesGalaxies2025}. To reduce the impact of cosmic variance, we have restricted our comparison to studies which probe survey areas of $\sim50~\text{arcmin}^2$ or larger. In addition, we plot number densities based on a purely spectroscopic sample by \citet{harikanePureSpectroscopicConstraints2024,harikaneJWSTALMAKeck2025} and measurements based on a spectroscopically-informed photometric sample by \citet{rojas-ruizBoRGJWSTSurveyAbundance2025}, where the photometric sample is corrected by the estimated contamination given the spectroscopically confirmed galaxies. The number densities from the literature are all plotted with what corresponds to 1$\sigma$ errors for a normal distribution.
Below, we compare with the literature in each dropout selection window.

Our number densities for the F090W-dropouts (median photometric redshift of $z\sim7.5$) are consistent within $1\sigma$ with the HST measurements by \citet{finkelsteinEVOLUTIONGALAXYRESTFRAME2015} and the JWST spectroscopic measurements by \citet{harikanePureSpectroscopicConstraints2024,harikaneJWSTALMAKeck2025} in all magnitude bins probed. Furthermore, at the bright end ($M_\mathrm{UV}<-21$), we are also consistent with the results from \citet{bouwensNewDeterminationsUV2021}, \citet{donnanEvolutionGalaxyUV2022}, \citet{adamsEPOCHSIIUltraviolet2024}, and the GO-2426 measurements from \citet{rojas-ruizBoRGJWSTSurveyAbundance2025}, but find some discrepancy with their GO-1747 measurements. At the faint end ($M_\mathrm{UV}>-19$), we are also consistent with \citet{bouwensNewDeterminationsUV2021} and \citet{adamsEPOCHSIIUltraviolet2024}. In the intermediate magnitude bins ($-21\leq M_\mathrm{UV}\leq-19$), where we have the highest number statistics, the number densities in this work are generally lower than most other works (up to $\sim0.5$ dex), however, most are consistent within the $2\sigma$ uncertainty range, except for the $M_\mathrm{UV}\approx-21$ measurement from \citet{adamsEPOCHSIIUltraviolet2024} and the \citet{willottSteepDeclineGalaxy2024} number densities, which are all above ours. Finally, comparing with the number densities estimated from BEACON DR1 \citep{morishitaBEACONJWSTNIRCam2025}, most magnitude bins exhibit consistent results within $1\sigma$; however, we find lower number densities from DR2 in the magnitude range $-22 \leq M_\mathrm{UV} \leq -20$, although still consistent within $2\sigma$.

For the F115W-dropouts ($z\sim10$), we measure number densities which are generally consistent within $1-2\sigma$ with the measurements by \citet{adamsEPOCHSIIUltraviolet2024}, \citet{willottSteepDeclineGalaxy2024} and BEACON DR1 \citep{morishitaBEACONJWSTNIRCam2025}, however our number densities fall up to $\sim0.5$ dex below other photometric selections in the literature, in bins where we have detections ($-21\leq M_\mathrm{UV} \leq -18$). Interestingly, we find excellent agreement with the spectroscopic measurements from \citet{harikanePureSpectroscopicConstraints2024,harikaneJWSTALMAKeck2025} at all magnitudes probed.

Finally, in the F150W-dropout selection, we find no robust galaxy candidates, so we obtain upper limits on the number density. Our $1\sigma$ upper limits are compatible with other observations from the literature \citep{mcleodGalaxyUVLuminosity2023, harikaneComprehensiveStudyGalaxies2023, finkelsteinCompleteCEERSEarly2024, donnanJWSTPRIMERNew2024, morishitaBEACONJWSTNIRCam2025, whitler$zGtrsim9$2025, weibelExploringCosmicDawn2026,francoPhysicalPropertiesGalaxies2025,whitler$zGtrsim9$2025}.

We tested the impact of varying, respectively, the width and the centres of the otherwise fixed $M_\mathrm{UV}$ bins, and found a negligible impact on the binned and parametric LFs for all three selections. This slightly changes the faint-end slope of the F090W-dropouts for both the Schechter- and DPL fit; however, this does not significantly alter our conclusions or the luminosity density (discussed below). We also tested the impact of potential contamination on our fitted LFs, assuming a 7.4\% contamination in our sample (corresponding to a 1$\sigma$ lower limit on the expected purity rate inferred from our 15 spectroscopic matches, see Sect.~\ref{sub:dropout-selection}), by masking out random sources in each dropout selection proportional to this contamination rate. The assumed 7.4\% contamination rate is comparable to the estimated interloper rate reported by \citet{leethochawalitUVRestframeBand2026} for their $z\sim8$ photometric sample adopting a broadly similar selection to our F090W-selection. From multiple iterations of sampling, we find that refitting the LFs leads to a negligible impact on the recovered UV LFs.

Overall, our inferred number densities from BEACON DR2 are consistent within $1\sigma$ with the purely spectroscopic measurements of \citet{harikanePureSpectroscopicConstraints2024,harikaneJWSTALMAKeck2025} in all three redshift bins. However, compared to recent photometric JWST determinations, our results generally lie on the lower side of the reported number densities. To assess this difference more quantitatively, we estimate how significant the difference in number counts is in BEACON compared with other JWST UV LF measurements.
Specifically, we compare the BEACON DR2 number counts for $M_\mathrm{UV}<-19.5$ sources -- the magnitude limit where most literature surveys are complete -- to the number of F115W- and F150W- dropouts expected assuming the Schechter LFs estimated by \citet{whitler$zGtrsim9$2025} in our survey volume (as estimated by our completeness simulations, see Sect.~\ref{sub:completeness}), accounting for both uncertainties in the LF parameters and Poisson statistics. 
For the F115W-dropout selection, we find a low probability ($p=0.0051$) of obtaining our observed counts or fewer assuming the \citet{whitler$zGtrsim9$2025} LF. This corresponds to a $\sim2.6\sigma$ discrepancy assuming a one-sided Gaussian equivalent, indicating that our number densities are systematically lower than their prediction, though with only moderate significance.
For the F150W-dropout selection, we obtain $p=0.0345$ (corresponding to a $\sim1.8\sigma$ difference), indicating no significant tension given the current number statistics. While we have not included cosmic variance in comparing to the JADES LFs, we expect it to reduce the significance of the F115W tension but not fully remove it, since the uncertainty is dominated by Poisson noise at $z\gtrsim 7$, due to the low number counts here. 

\subsection{The UV Luminosity Density and Comparison with Models}
\label{sub:LF_models}

We compute the UV luminosity density, $\rho_\mathrm{UV}$, as the luminosity-weighted integral of the fitted UV LFs, integrating down to a faint-end limit of $M_\mathrm{UV}=-17$:
\begin{equation}
    \rho_{UV} = \int_{-\infty}^{-17} L_\mathrm{UV} \cdot \phi(M) \; dM,
\end{equation}
where $L_\mathrm{UV}$ is the monochromatic luminosity at 1500\,\AA\ and $\phi(M)$ is the best-fit UV LF found in Sect.~\ref{sub:LF_fitting}. 

\begin{figure}
    \centering
    \includegraphics[width=\columnwidth]{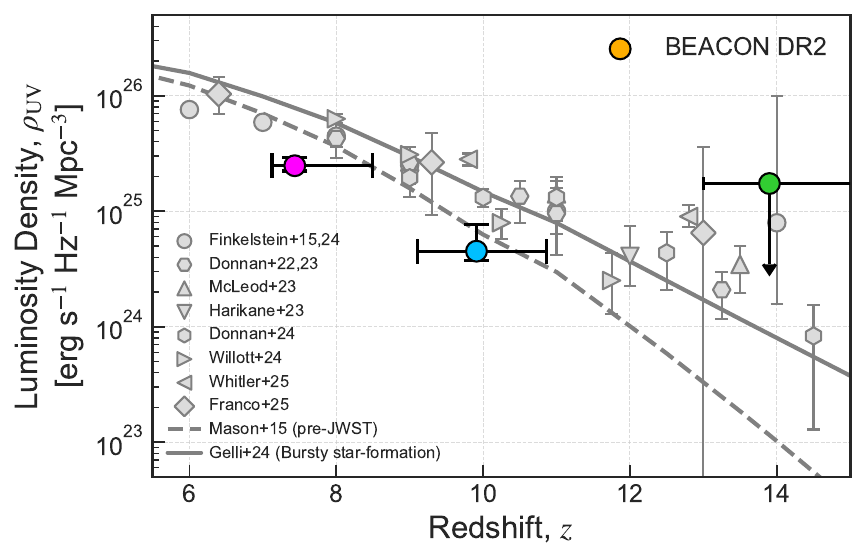}
    \caption{The cosmic UV luminosity density, $\rho_\mathrm{UV}$, resulting from integrating our best-fit Schechter function UV LFs down to $M_\mathrm{UV}=-17$ (coloured points). The errors on the luminosity density are estimated as the 16th and 84th percentiles of the $\rho_\mathrm{UV}$-values. The redshift error corresponds to the 16th and 84th percentiles of the best-fit \texttt{EAZY} redshift for the galaxy candidates included in the given sample (i.e. F090W- or F115W-dropouts). For the F150W-selection, where the luminosity function is unconstrained, we instead show an upper limit on $\rho_\mathrm{UV}$ derived from integrating the binned upper limits of the luminosity function over the range $-23 < M_\mathrm{UV} < -17$.
    For comparison, we show observational constraints from HST by \citet{finkelsteinEVOLUTIONGALAXYRESTFRAME2015} and from other JWST studies \citep{donnanEvolutionGalaxyUV2022, mcleodGalaxyUVLuminosity2023, harikaneComprehensiveStudyGalaxies2023,willottSteepDeclineGalaxy2024, finkelsteinCompleteCEERSEarly2024, donnanJWSTPRIMERNew2024, whitler$zGtrsim9$2025,francoPhysicalPropertiesGalaxies2025}. We also show predictions from theoretical models by \citet{masonGALAXYUVLUMINOSITY2015} and \citet{gelliImpactMassdependentStochasticity2024}}
    \label{fig:rhouv}
\end{figure}

\begin{figure*}
    \centering
    \includegraphics[width=\textwidth]{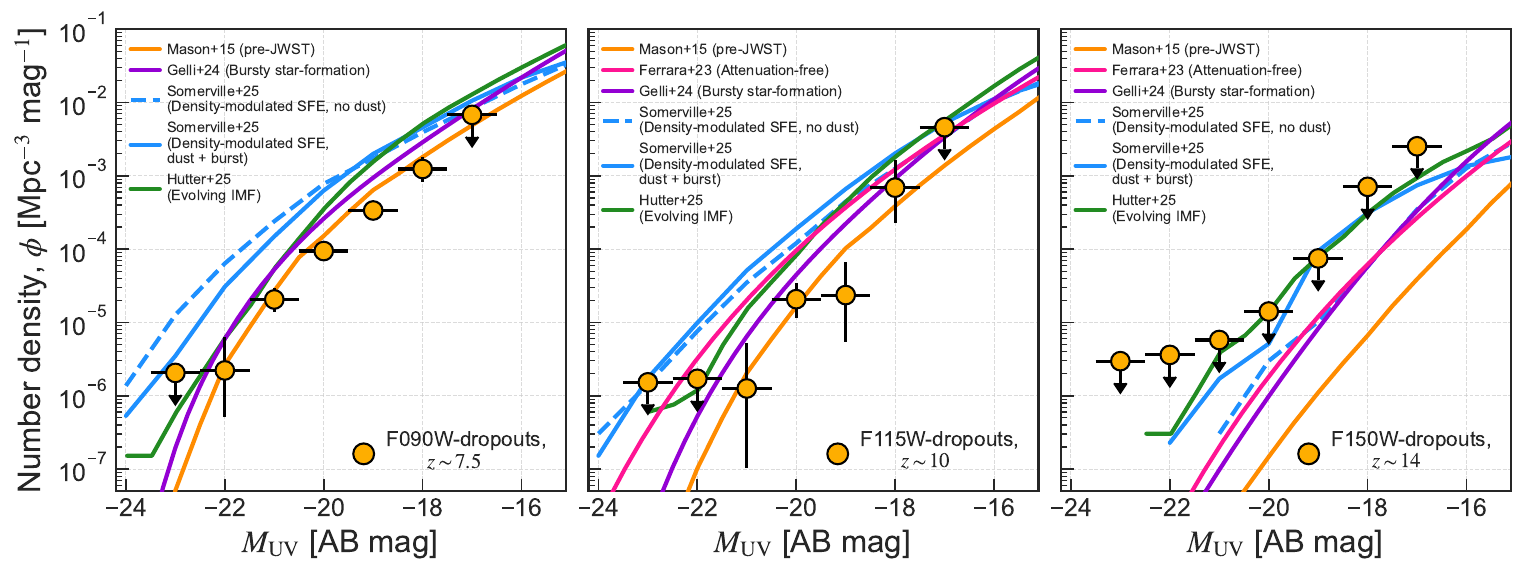}
    \caption{A comparison of the number densities measured from BEACON DR2 (orange scatter points) for, respectively, the F090W-dropouts ($z\sim7.5$), the F115W-dropouts ($z\sim10$), and the F150W-dropouts ($z\sim14$) with various theoretical models at similar redshifts (from left to right, $z=8$, $z=10$, and $z=14$). We plot an analytical pre-JWST model by \citet{masonGALAXYUVLUMINOSITY2015} (orange), implementing mass-dependent but redshift-independent SFE, an attenuation-free model by \citet{ferraraStunningAbundanceSuperearly2023} (pink), a model implementing halo mass-dependent stochasticity by \citet{gelliImpactMassdependentStochasticity2024} (violet), a density modulated SFE model by \citet{somervilleDensitymodulatedStarFormation2025} -- including a version with no dust (dashed blue), and a version incorporating dust and starbursts (solid blue), and a model with an evolving IMF by \citet{hutterASTRAEUSIndicationsTopheavy2025} (green).}
    \label{fig:uvlfs_models}
\end{figure*}

In Tab.~\ref{tab:uvlf_params}, we report the luminosity density for the F090W- and F115W-dropout selections using, respectively, the best-fit Schechter and DPL parameterisation of the UV LF. The lower and upper limits are computed from the 16th and 84th percentiles of the luminosity densities computed from the UV LF resulting from 1000 realisations of the MCMC parameter samples. The values obtained using the two parameterisations are similar and mutually consistent for both dropout selections. For the F150W-dropout selection, we instead compute an upper limit on the luminosity density. Following the approach of \citet{whitler$zGtrsim9$2025}, we take the luminosity-weighted integral of the binned values of the upper limits on the LF, i.e. between $M_\mathrm{UV}=-23$ and $M_\mathrm{UV}=-17$. We find an upper limit of $\rho_\mathrm{UV}(M_\mathrm{UV}<-17)=1.74\cdot 10^{25}$ erg s$^{-1}$ Hz$^{-1}$ Mpc$^{-3}$ for the F150W-dropout selection.

In Fig.~\ref{fig:rhouv}, we plot the luminosity densities derived from our Schechter UV LF fits, along with other observations from the literature \citep{finkelsteinEVOLUTIONGALAXYRESTFRAME2015, donnanEvolutionGalaxyUV2022, mcleodGalaxyUVLuminosity2023, harikaneComprehensiveStudyGalaxies2023,willottSteepDeclineGalaxy2024, finkelsteinCompleteCEERSEarly2024, donnanJWSTPRIMERNew2024, whitler$zGtrsim9$2025,francoPhysicalPropertiesGalaxies2025}. For comparison, we also show the pre-JWST model assuming redshift-independent star formation efficiency by \citet{masonGALAXYUVLUMINOSITY2015} and a model implementing halo mass-dependent stochasticity by \citet{gelliImpactMassdependentStochasticity2024}.

We see the luminosity density decreases with increasing redshift, as expected by models. Our luminosity densities derived from the F090W- and F115W-dropouts are lower than those reported from other works. At $z\sim7.5$ and $z\sim10$, our luminosity density is thus more consistent with the pre-JWST \citet{masonGALAXYUVLUMINOSITY2015} model. 
Our upper limit on the luminosity density at $z\sim14$ (derived from the F150W-dropouts) is consistent with both the pre-JWST model and the increased luminosity density reported in the literature.

Nonetheless, models which fit the UV luminosity density do not necessarily fit the observed shape of the UV LF, as they may, for example, have very steep faint-end slopes which increase the luminosity density. 
Thus, in Fig.~\ref{fig:uvlfs_models}, we plot our number densities in comparison to five classes of theoretical models: 
(i) an analytical pre-JWST model by \citet{masonGALAXYUVLUMINOSITY2015} implementing constant SFE and calibrated to reproduce the \citet{Bouwens2015b} $z\sim7$ HST UV LFs, (ii) an attenuation-free model by \citet{ferraraStunningAbundanceSuperearly2023}
(iii) a model implementing halo mass-dependent stochasticity by \citet{gelliImpactMassdependentStochasticity2024}, 
(iv) a density modulated SFE model by \citet{somervilleDensitymodulatedStarFormation2025} (including a version with no dust, and a version incorporating dust and starbursts), and (v) a model with an evolving IMF by \citet{hutterASTRAEUSIndicationsTopheavy2025}.

At $z\sim7.5$, our measurements are most compatible with the pre-JWST model by \citet{masonGALAXYUVLUMINOSITY2015}, lying below all other post-JWST models. However, at the intermediate magnitude bins ($-20\leq M_\mathrm{UV}\leq-18$), our number densities, while consistent within 1$\sigma$ with the model, are slightly below the \citet{masonGALAXYUVLUMINOSITY2015} model, resulting in a lower measured luminosity density than what the \citet{masonGALAXYUVLUMINOSITY2015} model predicts (Fig.~\ref{fig:rhouv}).
At $z\sim10$, our number densities are more uncertain due to the lower galaxy counts. While our measurements look most consistent with the \citet{masonGALAXYUVLUMINOSITY2015} model, they could also be compatible with the \citet{gelliImpactMassdependentStochasticity2024} model predicting higher number densities relative to pre-JWST predictions due to mass-dependent stochasticity in star formation.

At redshift $z>13$, we only have upper limits on the number density, making a distinction between theoretical models difficult. In fact, all theoretical models compared with are consistent with our constraints, and thus our results cannot rule out an excess of galaxies relative to pre-JWST models at this redshift.

\section{Characterisation of Overdensities}
\label{sec:overdensity}
Linking UV-bright high redshift galaxies to their underlying dark matter halos via clustering estimates also promises new insights into what drives the evolution in the UV luminosity function.
In this section, we leverage the large number of BEACON fields to identify candidate overdensities and investigate the environments of UV-bright galaxies.
We describe a framework for quantifying the probability that a field is overdense within a given redshift window (Sect.~\ref{sub:overdense_prob}), and use this to assess whether the brightest galaxies in our sample are more likely to reside in overdense fields (Sect.~\ref{sub:bright_overdense}).

\subsection{Quantifying the Probability that a Field is Overdense}
\label{sub:overdense_prob}

Following the approach of \citet{trentiOverdensitiesYdropoutGalaxies2012}, we aim to quantify the probability of detecting $N_\mathrm{obs}$ sources in a single pointing, within a certain redshift window, and subsequently use this to assess the statistical significance of a field being overdense.

Assuming Poisson statistics and no clustering, the probability of observing $N_\mathrm{obs}$ or more sources in one field where $N_\mathrm{exp}$ sources are expected is given by:
\begin{equation}
     P(N\geq N_\mathrm{obs} \mid N_\mathrm{exp})=
     1 -
     e^{-N_\mathrm{exp}} \sum_{j=0}^{\lfloor N_\mathrm{obs}-1 \rfloor} \frac{N_\mathrm{exp}^j}{j!},
     \label{eq:P_geqNobs}
\end{equation}
where the last term represents the cumulative distribution function (CDF) of a Poisson distribution with mean $N_\mathrm{exp}$.
The lower the value of $P(N\geq N_\mathrm{obs} \mid N_\mathrm{exp})$, the more unlikely it is to observe $N_\mathrm{obs}$ galaxies if the counts were Poisson-distributed, and thus the more statistically significant it is that a field is overdense.
When $N_\mathrm{exp}\lesssim 1$ (which is the case in most BEACON fields), Poisson uncertainty dominates over cosmic variance, and Eq.~\ref{eq:P_geqNobs} provides a good description of the significance of overdensities.

We calculate this Poisson probability for each field in the following way.
The observed number of objects in each field, $N_\mathrm{obs}$, is the number of sources brighter than $M_\mathrm{UV}=-17$ in our sample, with photometric redshifts within the investigated redshift window, $z_\mathrm{low}<z<z_\mathrm{up}$.
Given the high purity of our selection (Sect.~\ref{sub:dropout-selection}), we do not account for contamination in $N_\mathrm{obs}$.
The expected number of galaxies in each field, $N_\mathrm{exp}$ is computed by the following. We integrate the global UV LF by interpolating our Schechter function best-fits onto a grid of redshift between $z\sim7\text{ -- }10$, $\phi_{\rm{interp}}(M,z)$, while accounting for the completeness function for each field (Sect.~\ref{sub:completeness}):
\begin{equation}
    N_{\rm{exp}} = \int_{-\infty}^{-17} 
                    \left[ \int_{z_{\rm{low}}}^{z_{\rm{up}}} 
                    \phi_{\rm{interp}}(M,z) P(M,z) \frac{dV}{dz} dz
                    \right] dM.
    \label{eq:Nexp}
\end{equation}

We linearly interpolate the UV LF in log space between the two best-fit Schechter curves (assuming the parameters in Tab.~\ref{tab:uvlf_params}) that were fitted to the number densities of, respectively, the F090W-dropouts (assuming $z_{\rm{median}}=7.47$ based on the median of the photometric redshifts in that selection) and the F115W-dropouts ($z_{\rm{median}}=10.14$), enabling us to estimate the LF at intermediate redshifts.
Since the F115W-dropout selection includes candidates with photometric redshift up to $z_\mathrm{phot}\sim12$, we fix the LF used in Eq.~\ref{eq:Nexp} to that of the F115W-dropouts for redshifts above $z_{\rm{median}}=10.14$.
We note that our integration limit, $M_\mathrm{UV}<-17$, is fairly arbitrary since many fields are incomplete at these magnitudes. However, this is accounted for by the completeness function, $P(M,z)$, in Eq.~\ref{eq:Nexp}. The completeness correction implies that galaxies fainter than $M_\mathrm{UV}\sim-19$ contribute negligibly to $N_\mathrm{exp}$. Adopting a shallower limit does not change our conclusions, since $N_\mathrm{obs}$ is defined using the same magnitude threshold.

\subsection{Are Fields Hosting the Brightest Galaxies More Likely to Be Overdense?}
\label{sub:bright_overdense}

\begin{figure*}
    \centering
    \includegraphics[width=\textwidth]{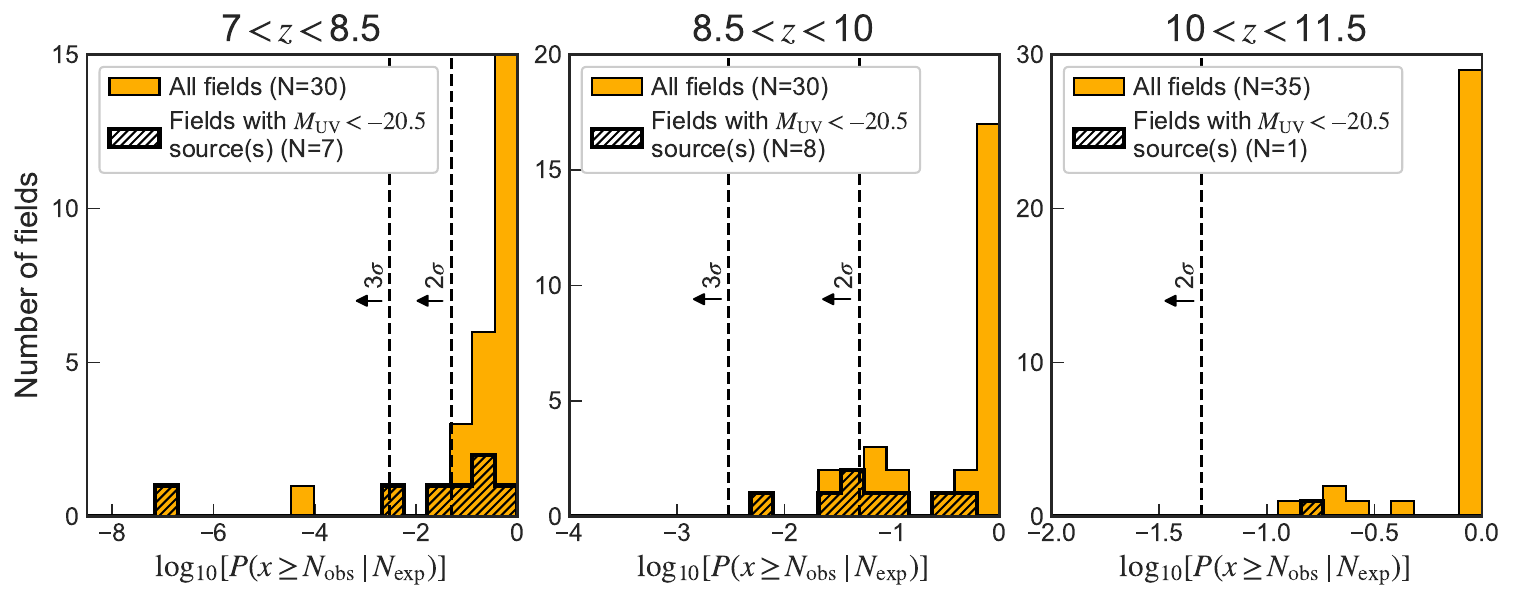}
    \caption{Distributions of the probability of observing $N_\mathrm{obs}$ sources or more in a field where $N_\mathrm{exp}$ sources are expected, measured for, respectively, all fields (orange) and fields hosting at least one bright $M_\mathrm{UV}<-20.5$ source (black, hatched), in three redshift bins: $7<z<8.5$, $8.5<z<10$, and $10<z<11.5$ (left to right panel). The lower the value of $P(N\geq N_\mathrm{obs} \mid N_\mathrm{exp})$, the more unlikely it is that we observed $N_\mathrm{obs}$ galaxies, and thus the more statistically significant it is that a field is overdense. The dashed lines mark regions corresponding to Poisson probabilities of $P(N\geq N_\mathrm{obs} \mid N_\mathrm{exp}) < 0.003$ and $<0.05$, equivalent to the one-sided $3\sigma$ and $2\sigma$ tail probabilities of a Gaussian distribution. Throughout the paper we refer to these thresholds as the $3\sigma$ and $2\sigma$ overdensity levels.}
    \label{fig:overdense_sigs}
\end{figure*}

We now aim to characterise the local densities of fields hosting the brightest galaxies in our sample, and quantify whether these are more overdense than the average field. We compute the distribution of $P(N\geq N_\mathrm{obs} \mid N_\mathrm{exp})$ (Eq.~\ref{eq:P_geqNobs}) in two groups: (1) all fields in our sample and (2) only fields hosting at least one $M_\mathrm{UV}<-20.5$ source. 
We conduct our analysis in $\Delta z=1.5$ redshift windows in the range $z=7\text{ -- }11.5$, where we mostly have sufficient dropouts in each field to estimate the overdensity without being dominated by Poisson noise. We sort our galaxy candidates into redshift bins based on the peak of their $p(z)$ distribution.

The resulting distributions are displayed in Fig.~\ref{fig:overdense_sigs} for the three redshift windows: $7<z<8.5$, $8.5<z<10$ (using the completeness function that employs the F090W-dropouts selection), and $10<z<11.5$ (using the F115W-dropouts completeness function).
The dashed lines mark the regions where the Poisson probability satisfies 
$P(N\geq N_\mathrm{obs} \mid N_\mathrm{exp}) < 0.003$ and $<0.05$. These correspond to the one-sided tail probabilities of $3\sigma$ and $2\sigma$ deviations for a Gaussian distribution. Although our significance is derived from Poisson statistics, for simplicity we adopt the equivalent Gaussian terminology and refer to these thresholds as the $3\sigma$ and $2\sigma$ significance levels throughout the remainder of this paper. Thus, if a field exhibits a probability $P(N\geq N_\mathrm{obs} \mid N_\mathrm{exp})$ within these regions, we consider the field overdense at the corresponding significance level.

At the $3\sigma$ level, we find two fields with candidate overdensities in the $7<z<8.5$ redshift window, one of which hosts at least one bright galaxy. 
Hosting two $M_\mathrm{UV}<-20.5$ sources ($M_\mathrm{UV}=-21.71^{+0.07}_{-0.06}$ and $M_\mathrm{UV}=-20.79^{+0.10}_{-0.11}$), we identify the field: \texttt{beacon\_1420+5252} (with $N_\mathrm{obs}=11$ and $N_\mathrm{exp}=1.28$). The field \texttt{beacon\_0015-3034} (with $N_\mathrm{obs}=9$ and $N_\mathrm{exp}=1.60$) is also overdense at the $3\sigma$ level, but does not host any $M_\mathrm{UV}<-20.5$ source. The brightest source in this field is $M_\mathrm{UV}=-20.12^{+0.07}_{-0.07}$. Additionally, the field \texttt{beacon\_2325-1203} (with $N_\mathrm{obs}=7$ and $N_\mathrm{exp}=1.65$), hosting an $M_\mathrm{UV}=-20.95^{+0.09}_{-0.07}$ source, is found to be overdense at the $2.7\sigma$ significance level.

The overdensities we identify are unlikely to be false positives arising from multiple testing, and are robust given our estimated contamination rate.
Since we evaluate overdensities across multiple field–redshift window combinations, some spurious overdensities are expected by chance; the quoted significances are local (single-trial) Poisson probabilities.
Considering all 95 field--window combinations analysed here and assuming them to be independent trials, the expected number of spurious overdensities is approximately $N_{\rm trial} \cdot p_{\rm threshold}$, corresponding to $\sim0.3$ and $\sim4.8$ false positives at the adopted $3\sigma$ ($P<0.003$) and $2\sigma$ ($P<0.05$) thresholds, respectively. 
Since most of these field--window combinations are in reality not independent, owing to e.g. some overlap in fields between redshift bins and adjacent redshift windows, these values are conservative upper limits.
Thus our identified two $>3\sigma$ overdensities in the $7<z<8.5$ window exceed the expected $<0.3$ spurious detections, and are thus very unlikely to be spurious. We also find that the two $>3\sigma$ overdense fields are robust against potential contamination in our sample. If we compute $N_\mathrm{exp}$ (Eq.~\ref{eq:Nexp}) from LFs assuming a $7.4\%$ contamination rate (see Sect.~\ref{sub:LF_results}) and assume one source in each field is a contaminant, both fields will remain $>3\sigma$ overdensities. We note, given the small $N_\mathrm{obs}$, even one contaminant would be more than the $1\sigma$ lower limit on the expected purity rate.

To understand whether our $>2.5\sigma$ overdense fields represent physical associations of galaxies (i.e. on $<$ tens of cMpc scales rather than the $\sim400$\,cMpc scales of our redshift windows), we analyse the redshift posterior distributions, $p(z)$, in each of the three fields. In all three fields, we identify overdensities of candidates at $>2.5\sigma$ significance within narrower $\Delta z=0.6$ windows, where the galaxies are more than $50\%$ likely (ranging between 58 -- 100\%, with a median of $\sim90\%$) to be separated within $\Delta z \leq 0.6$ based on their respective $p(z)$. Thus, we consider these photometric overdensities are likely to be physically associated galaxies.

Our most overdense field, \texttt{beacon\_1420+5252}, overlaps with the footprint of the EGS field, which is known to be one of the most overdense fields at $z\sim7-9$ across all extragalactic legacy fields \citep[e.g.][]{leonova_prevalence_2022,chenImpactGalaxyOverdensities2026,whitlerDeepJWSTSpectroscopy2026}. The EGS field also shows one of the highest Lyman-alpha emitter fraction across legacy fields at $z>7$ (see \citealp{tangJWSTNIRSpecObservations2024,Napolitano2024}, c.f. \citealp{Jones2025}) and contains several spectroscopic overdensities, and thus likely hosts some of the largest ionized bubbles known at these redshifts \citep{Jung2022,tangJWSTNIRSpecSpectroscopy2023a,tangJWSTNIRSpecObservations2024,Napolitano2024,chenImpactGalaxyOverdensities2026}. This field is identified as the most overdense of all 30 BEACON fields where we can perform F090W selections, and thus adds further evidence that EGS is an exceptional location for understanding the link between structure formation and reionisation.
The field \texttt{beacon\_0015-3034} is located $\sim0.2\,\text{deg}$ from both the centre of a proto-cluster spectroscopically confirmed at $z=7.88$ in the Abell-2744 field \citep[e.g.][]{morishitaEarlyResultsGLASSJWST2023} and a potential overdensity around the $z=7.04$ LRD Abell2744-QSO1 \citep{Tang2026}, corresponding to a physical separation of $\sim(25)35$\,cMpc at $z=(7)8$, and therefore might hint at a potential large-scale clustering at $z\sim7-8.5$ in this region, though more complete spectroscopy will be required to confirm the redshift of the overdensity. Lastly, the field \texttt{beacon\_2325-1203} is a novel extra-galactic field with BEACON providing the first NIRCam coverage of this region. The physical properties of the galaxies in these three fields are investigated in \citet{zhangBEACONJWSTNIRCam2026}.

We now quantify whether fields hosting bright galaxies are more likely to be overdense compared to a random field. For the two subsamples of fields (all fields that probe the given redshift, and all of those fields that host at least one $M_\mathrm{UV}<-20.5$ source), we compute the fraction of fields that are overdense with, respectively, a $2\sigma$ and $3\sigma$ significance in each redshift window.
At $7<z<8.5$, the fraction of fields that are overdense at the $3\sigma$ ($2\sigma$) level increases from 7 to 14\% (13 to 43\%) across all fields versus only fields hosting bright galaxies.
At $8.5<z<10$, we find no evidence of fields that are overdense with $3\sigma$ significance. However, at the $2\sigma$ level, we observe the same trend as in the previous redshift bin: the fraction of overdense fields rises from 17 to 50\% when restricting the sample to fields hosting bright sources.
At the highest redshifts investigated here, $10<z<11.5$, we detect fewer sources, and are thus more dominated by Poisson noise (when $N_\mathrm{obs} = 1$). Consequently, we observe no fields that are likely to be overdense at either the $2\sigma$ or $3\sigma$ level.
Nonetheless, we find that the field hosting the brightest galaxy at $10<z<11.5$ is more likely to be overdense compared to the full sample of fields, albeit with lower significance.

Overall, our sample demonstrates significant clustering of $M_\mathrm{UV} < -20.5$ galaxies at $z>7$, finding that fields hosting such bright galaxies are almost three times ($2.1-3.2\times$) as likely to be overdense compared to our full sample of fields. We discuss the implications of this in Sect.~\ref{sub:disc_bright_environments}. We note that while sources close to the edges of the redshift windows could in principle affect the overdensity probabilities, we here find that fields hosting such bright galaxies are still $2.1-3.2\times$ more likely to be overdense if we repeat the analysis computing $N_\mathrm{obs}$ as the number of sources in each field that satisfies the $M_\mathrm{UV}$ limit and which have an integrated $p(z)$ inside the given redshift interval of 50\% or more.

\section{Discussion}
\label{sec:disc}
In Sect.~\ref{sub:steep_UVLF_evolution}, we discuss the evolution of the UV luminosity function found in this work in comparison with the literature. In Sect.~\ref{sub:z13_abundance}, we discuss the implications of our results for constraining the number densities of $z>13$ galaxies. Finally, in Sect.~\ref{sub:disc_bright_environments}, we discuss the implications of our finding that UV-bright galaxies are more likely to reside in overdense fields.

\subsection{The Evolution of the UV Luminosity Function}
\label{sub:steep_UVLF_evolution}

In Sect.~\ref{sec:uvlfs}, we presented the UV luminosity function from the BEACON DR2 dataset in three different redshift bins, $z\sim7.5,10,$ and $z>13$, respectively.
While the redshift evolution of our LFs is qualitatively similar to previous work, we find lower number densities at $-21 \lesssim M_\mathrm{UV} \lesssim - 19$ at $z\sim7.5$, and lower number densities at $z\sim10$ compared to many previous works, though most observations are still consistent within $1-2\sigma$, as we will discuss more below.
At $z\sim10$, our UV LF is consistent with pre-JWST model predictions (see Sect.~\ref{sub:LF_models}).
This is a similar finding to \citet{adamsEPOCHSIIUltraviolet2024} and \citet{willottSteepDeclineGalaxy2024}.
Finally, we find no robust $z>13$ galaxy candidates, but our upper limits on the number density do not rule out an excess of sources relative to pre-JWST predictions (see Fig.~\ref{fig:uvlfs_models}).

We now seek to explore potential explanations for differences in number densities in BEACON DR2 relative to other studies at $z\sim7.5$ and $z\sim10$.
One effect which may be important is cosmic variance, which should be minimised in BEACON due to our large number of independent pointings. Some of the studies we compare with are indeed based on only 1-2 independent fields \citep[e.g.][]{donnanAbundance10Galaxy2023, finkelsteinCompleteCEERSEarly2024, whitler$zGtrsim9$2025}, which could explain some of the discrepancies, especially at higher redshift, e.g. if those observed sightlines are more overdense than the cosmic average. However, we also compare with studies that combine data from $>8$ independent sightlines, including the Cycle 1 pure parallel program PANORAMIC \citep[e.g.][]{adamsEPOCHSIIUltraviolet2024, weibelExploringCosmicDawn2026, mcleodGalaxyUVLuminosity2023}, which are thus less susceptible to cosmic variance, and still find the BEACON LFs lie below most of these at $z\sim10$.
Thus, while cosmic variance may explain a component, it can not alone account for the differences we are seeing.

Another potential explanation for the discrepancies between our measured number densities and others is the selection we apply in this study, which may be more conservative than others in the literature. 
We require robust non-detections blueward of the break with S/N criteria of $<2\sigma$, which is more stringent compared to e.g. \citet{adamsEPOCHSIIUltraviolet2024} who apply a $<3\sigma$ cut.
Our use of the F090W filter also improves the robustness of F115W- and F150W-dropouts, compared to studies that have no comparable $\sim0.8\text{--}1\, \mu \mathrm{m}$ imaging in at least some of their survey area \citep[e.g.][]{donnanEvolutionGalaxyUV2022,donnanAbundance10Galaxy2023,mcleodGalaxyUVLuminosity2023,weibelExploringCosmicDawn2026}. 
While some legacy fields have HST/F814W imaging (e.g. \citealp[]{finkelsteinCompleteCEERSEarly2024,francoPhysicalPropertiesGalaxies2025}), this typically has lower sensitivity than our F090W imaging.
Our selection also incorporates strong constraints on the redshift probability distribution. As described in Sect.~\ref{sub:dropout-selection}, we require that at least 80\% of the integrated $p(z)$ lies above a redshift threshold, $z_{\rm min}$, that increases with each dropout category ($z_{\rm min}=6,8,10$ for respectively the F090W-, F115W-, and F150W-dropout selections). This tiered $p(z)$ requirement is more restrictive than other recent UV LF studies \citep[e.g.][]{donnanEvolutionGalaxyUV2022,harikaneComprehensiveStudyGalaxies2023,weibelExploringCosmicDawn2026}. Others adopt a single $p(z)$ threshold (typically tied to an initial $z_{\rm min}\sim6-7$), and then use the best-fitting photometric redshift to sort galaxies into redshift subsamples \citep[e.g.][]{ willottSteepDeclineGalaxy2024, finkelsteinCompleteCEERSEarly2024}. Studies that implement a redshift-dependent probability cut similar to ours often adopt a lower probability requirement of 50 -- 60\% above their $z_{\rm min}$ thresholds \citep[e.g.][]{whitler$zGtrsim9$2025, francoPhysicalPropertiesGalaxies2025}. 

These different choices in S/N filtering and $p(z)$ selection directly affect the fraction of borderline sources admitted into high-$z$ samples.
As such, BEACON’s more conservative strategy, designed to maximise purity, is expected to yield systematically smaller candidate samples. 
To illustrate this, relaxing our $p(z)$ requirements to $p(z_\mathrm{phot}>z_{\rm min}) > 70\%$ and $\chi^2_{\rm{high-z}} - \chi^2_{\rm{low-z}} < -3$ increases the number of candidates from 164 to 193, recovering an additional spectroscopically confirmed source from the DJA v4.4 catalogue that did not quite satisfy our fiducial $p(z)$-requirements (see Sect.~\ref{sub:dropout-selection}). However, this increase is accompanied by a corresponding rise in completeness, leading to a larger effective survey volume, and thus negligible differences in the measured number densities.
However, this is contingent upon the properties of any missed sources being well-captured by the JAGUAR templates \citep{williamsJWSTExtragalacticMock2018} that we use in our completeness simulations (see Sect.~\ref{sub:completeness}). We note the JAGUAR catalogue has an imposed limit for the UV slope, $\beta_\mathrm{min} \geq -2.6$, yet UV slopes as blue as $\beta<-2.8$ have been reported \citep[e.g.,][]{toppingUVContinuumSlopes2024}. As very blue sources become more common at fainter magnitudes, this suggests that we might be overestimating the completeness at fainter magnitudes; however, we note the fraction of $\beta_\mathrm{min} < -2.6$ is only a few percent in current samples \citep{toppingUVContinuumSlopes2024,Cullen2024,Morales2024}, so this is unlikely to lead to a significant underestimate of the inferred LF.

While a more conservative selection reduces the total number of candidates in the final sample, it also ensures high confidence in the sources included. Indeed, we find no low-z interlopers and thus an observed purity rate of 100\% among the sources that passed our selection criteria that have also been spectroscopically confirmed (see Fig.~\ref{fig:zspec}).
The observed purity rate, when reported, in other studies we compare to is lower, $\sim80-90\%$ \citep{harikanePureSpectroscopicConstraints2024,adamsEPOCHSIIUltraviolet2024,finkelsteinCompleteCEERSEarly2024,weibelExploringCosmicDawn2026,whitler$zGtrsim9$2025}, though we note \citet{willottSteepDeclineGalaxy2024}, who, similarly to us, find lower number densities, also report no low-z interlopers.
Contamination is not commonly accounted for, but, especially at redshifts $z>10$, where there is a lower number of sources, it is plausible that the contamination rate has a non-negligible impact on the inferred UV LFs. However, given the high purity of our sample, we do not observe a significant impact of potential contamination on the recovered LFs in this study (for more details see Sect.~\ref{sub:LF_results}).

In fact, when we compare our number densities with those found from a purely spectroscopic sample by \citet{harikanePureSpectroscopicConstraints2024,harikaneJWSTALMAKeck2025}, we find excellent agreement at redshifts $z\sim7.5$ and $z\sim10$. At $z\sim14$, the measurement from \citet{harikanePureSpectroscopicConstraints2024,harikaneJWSTALMAKeck2025} is above our upper limit, but their 1$\sigma$ error remains consistent with our constraints. We note that the $z>13$ sources in the \citet{harikanePureSpectroscopicConstraints2024,harikaneJWSTALMAKeck2025} spectroscopic sample were only identified in GOODS-S from JADES \citep{carnianiSpectroscopicConfirmationTwo2024,witstokWitnessingOnsetReionization2025}, which may be overdense at $z\sim14$ \citep{Robertson2024}.
The excellent agreement between BEACON LFs and the spectroscopic LFs by \citet{harikanePureSpectroscopicConstraints2024,harikaneJWSTALMAKeck2025} suggests that selection effects likely drive the differences between our measured number densities and others in the literature, with our sample most likely exhibiting high purity but consequently less inclusive than others.

Prospects for accurately measuring the high redshift UV luminosity function are promising, thanks to the sensitivity of JWST/NIRSpec. Dedicated prism surveys, e.g. CAPERS (GO-6368, PI: M. Dickinson) \citep{kokorevCAPERSObservationsTwo2025, donnanVeryBrightVery2025} and MoM (GO-5224, PI: P. Oesch) \citep{naiduCosmicMiracleRemarkably2026} have begun to confirm $z>9$ photometric candidates. These will be crucial for optimising photometric selections, though they are still relatively shallow (1.58 -- 4.74 hr depths). In the future, deep prism and grating spectra will enable us to establish the robustness of $z>9$ selections, particularly for faint sources. 

Finally, our results imply it would also be valuable to carry out more dedicated comparisons of the UV LF as measured by HST and JWST at $M_\mathrm{UV} \gtrsim -21$ at $z\sim7.5$, to properly understand the shape of the UV LF at this redshift. 
As pointed out by \citet{morishitaPhysicalCharacterizationEarly2023}, none of the spectroscopically confirmed $z>7$ sources in the first JWST field SMACS-0723 had been successfully selected by a previous study using HST data. This is likely a combination of the accuracy of photometric redshifts from HST's limited wavelength coverage and the impact of morphology. Recent JWST observations report a variety of morphologies even at the cosmic frontier and in UV faint galaxies \citep{harikaneJWSTALMAKeck2025,carnianiSpectroscopicConfirmationTwo2024}. In shallow, few-band imaging, detection incompleteness is expected to be more significant for galaxies with extended morphologies, which may introduce differences between HST and JWST-derived UV LFs.

\subsection{The Abundance of $z>13$ Galaxies}
\label{sub:z13_abundance}

JWST has pushed our horizon of galaxy detection to $z>11$ for the first time, with $z\sim14$ being the current spectroscopic frontier \citep{carnianiSpectroscopicConfirmationTwo2024,naiduCosmicMiracleRemarkably2026}.
While BEACON DR2 has expanded the search for $z>13$ sources over 19 fields, we find no robust $z>13$ galaxy candidates, demonstrating the importance of cosmic variance at these redshifts. 
Nonetheless, the large area probed in BEACON DR2 (216 sq. arcmin for this selection, see Tab.~\ref{tab:candidates}) enables us to place strong constraints on the number densities at these redshifts. As discussed in Sect.~\ref{sub:LF_models}, our results do not rule out the excess of luminous sources at $z>13$ that has been suggested by other observations \citep[e.g.][]{mcleodGalaxyUVLuminosity2023,finkelsteinCompleteCEERSEarly2024,whitler$zGtrsim9$2025,francoPhysicalPropertiesGalaxies2025}. While we find no robust F150W-dropout galaxies across our 19 BEACON fields probing these redshifts, we find no strong tension when comparing our observed number count of zero with the predicted counts in our survey volume based on the \citet{whitler$zGtrsim9$2025} F150W-dropouts LF (see Sect.~\ref{sub:LF_results}).

Still, three $z\sim14$ galaxies are now spectroscopically confirmed \citep{carnianiSpectroscopicConfirmationTwo2024,naiduCosmicMiracleRemarkably2026}.
These show very different morphologies, spectra, and physical properties, implying potentially a range of mechanisms or an evolutionary effect driving their extreme luminosities \citep[see also][]{roberts-borsaniJWSTSpectroscopicInsights2026,tangJWSTSpectroscopicProperties2026}.
Understanding what drives the abundance of bright $z>14$ galaxies will ultimately require larger samples to explore their detailed physical properties, as well as deeper follow-up to constrain their environments (see Sect.~\ref{sub:disc_bright_environments} below).
The prospects for establishing a larger sample for spectroscopic follow-up will grow with increased survey area across the JWST mission.
Given the predicted cosmic variance in the UV LF at $z>7$, pure parallel observations provide an efficient way to build larger samples of bright $z\sim14$ galaxies. Adopting the F150W-dropout UV LF from \citet{whitler$zGtrsim9$2025}, we calculate that $>100$ NIRCam pointings would be required to obtain a sample of $>10$ galaxies with $M_{UV} < -20$ at $z \gtrsim 14$. Progress should be expected as JWST continues to build a legacy of deep multi-band imaging.

Interestingly, theoretical models invoked to explain the overabundance of $z\sim14$ galaxies begin to diverge significantly in their predicted UV LF towards higher redshifts \citep[e.g. see compilations of models reported by][]{Castellano25, perez-gonzalezRiseGalacticEmpire2025}, making accurate measurements from larger samples essential for discriminating between competing physical scenarios.
However, the search for galaxies at higher redshift becomes more difficult as the number of detection filters decreases. The selection of $z\gtrsim17$ galaxies (F200W-dropouts) is severely subject to confusion by cosmic rays. In particular, cosmic rays hitting only red detectors but not blue can produce artificial breaks, e.g. between the F277W and F200W band, which are challenging to distinguish visually given the expected size of the source: the pixel size of the NIRCam red channel is $2\times$ larger than that of the blue channel. While dithering can help reduce the impact of cosmic rays, the constraints of pure-parallel programs may leave some observations with insufficient observing slots to enable the number of dithers required for reducing the effects of cosmic rays to a reasonable level that enables robust F200W-dropout selection. The addition of medium bands in the short and long-wavelength channels can improve F200W selection by more finely sampling the break, improving photometric redshift estimates, reducing cosmic ray contaminants, and distinguishing low-$z$ emission line contaminants \citep{eisensteinJADESOriginsField2025}.

\subsection{The Environments of UV-Bright Galaxies at $z>7$}
\label{sub:disc_bright_environments}

One of the most promising avenues to understand what drives the UV LF evolution at high redshift is to establish what mass range of dark matter halos the brightest galaxies reside in.
Different galaxy formation models predict significantly different relationships between halo mass and galaxy luminosity, which can be tested with galaxy clustering estimates \citep[e.g.][]{Ren2018,Mirocha2020b,munozBreakingDegeneraciesFirst2023,gelliImpactMassdependentStochasticity2024}.
Most models for the UV LF excess observed by JWST fall into two classes: either galaxies are brighter at fixed halo mass due to, e.g. an increase in star formation efficiency \citep{dekelEfficientFormationMassive2023,somervilleDensitymodulatedStarFormation2025}, lower dust obscuration \citep{ferraraStunningAbundanceSuperearly2023}, and/or a more top-heavy IMF \citep{inayoshiLowerBoundStar2022, cuetoASTRAEUSIXImpact2024,hutterASTRAEUSIndicationsTopheavy2025}; or very stochastic star formation temporarily upscatters galaxies in low-mass halos to high UV luminosities \citep{masonBrightestGalaxiesCosmic2023,Sun2023,shenImpactUVVariability2023,gelliImpactMassdependentStochasticity2024}.
If galaxies are on average brighter, UV-bright galaxies should still reside in the most massive halos at a given redshift, tracing high-density peaks of the matter field, and thus these galaxies should be strongly clustered, while if stochasticity becomes more important at high redshift, UV-bright galaxies are hosted in a broader range of halo masses, and thus less clustered.

Wide-area surveys like BEACON have the unique ability to identify bright $z\sim7-14$ galaxies and characterise their environments. Crucially, JWST has the sensitivity to detect faint $M_\mathrm{UV}\sim -17$ galaxies in the surroundings of bright sources, and the wavelength coverage for robust photometric selection, which are beyond the sensitivities and wavelength range of wide-area near-infrared telescopes like Euclid and Roman \citep{euclidcollaborationEuclidPreparationEuclid2022,euclidcollaborationEuclidOverviewEuclid2025,committeeRomanObservationsTime2025}.
In Sect.~\ref{sec:overdensity}, we demonstrated that fields hosting $M_\mathrm{UV}<-20.5$ galaxies at $z>7$ are significantly more likely to be overdense compared to the full survey, implying they trace massive halos.

To translate the observed environments of bright galaxies into constraints on their host halo masses, one can use measurements of galaxy clustering \citep[see e.g.,][for a review]{Wechsler2018}. The strength of clustering reflects how strongly galaxies trace the underlying dark matter density field, and thus provides an estimate of the galaxy bias and the typical mass of the halos they inhabit. Clustering is traditionally quantified through the two-point correlation function (2PCF), but reliably estimating this requires both large contiguous areas and deep multi-band imaging, though initial estimates of correlation functions at $z>7$ have been obtained with JWST in small fields \citep{dalmassoGalaxyClusteringCosmic2024,shuntovConstraintsEarlyUniverse2025,dalmassoAcceleratedEvolutionGalaxy2026} and shallower wide-area fields \citep[][]{paquereauTracingGalaxyhaloConnection2025}. An alternative approach is the count-in-cells method proposed by \citet{Robertson2010}, which estimates the galaxy bias, $b=\delta_{\mathrm{gal}}/\delta_{\mathrm{DM}}$, from the variance in galaxy number counts across many independent sightlines. This method has been successfully applied at $z\sim2$ with HST using 141 independent fields \citep{cameronObservationalDeterminationGalaxy2019}. Recently, \citet{weibelExploringCosmicDawn2025a} carried out a similar analysis at $z\sim10$ with JWST using 34 independent NIRCam fields. That work provided the first empirical constraints on cosmic variance at $z\sim10$, though they find that precise estimates of galaxy bias are not yet feasible with current JWST datasets, as the number of independent fields remains too small and the NIRCam field of view does not probe sufficiently large scales to measure linear clustering.

Given these limitations, we here adopt a simpler approach and compare the observed overdensity statistics of bright galaxies to simulated galaxy catalogues, which capture non-linear clustering. Specifically, we generate density fields in $(1\,\mathrm{cGpc})^3$ boxes with a resolution of 1\,cMpc at $z=7.5, 9.5$, and $10.5$ using the semi-numerical code \texttt{21cmFAST} \citep{Mesinger2011,murray21cmFASTV3Pythonintegrated2020}. We sample halos from the density field using excursion set theory as implemented by \citet{daviesEfficientSimulationDiscrete2025}, which has been demonstrated to reproduce halo statistics from N-body simulations down to 1\,cMpc. We then apply the UV luminosity to halo mass relation from \citet{masonGALAXYUVLUMINOSITY2015} (assuming no scatter in the relation) as a function of redshift to each halo in the simulation to generate a 3D galaxy catalogue. This model has been shown to reproduce $z\sim0-10$ UV LFs and the $z\sim4-7$ luminosity - halo mass relations inferred from clustering observations \citep{Harikane2017}.

To emulate the volumes we probe in our analysis in Sect.~\ref{sec:overdensity}, we extract 10,000 mock fields from the simulation box. Each field has an angular area corresponding to the field of view of NIRCam and a depth corresponding to $\Delta z = 1.5$, with centre redshifts of $z_\mathrm{mid}=7.5,9.5$, and $10.5$. The centroids of the mock fields are uniformly sampled from the simulation, ensuring that the sampled volume does not extend outside the full simulation box to avoid edge effects. 
Within each mock field, we count the number of galaxies, $N_\mathrm{obs}$, brighter than the limit of $M_\mathrm{UV}<-17$ that we adopted for the equivalent analysis of the observations. The expected number of sources in each mock field is found from the density of $M_\mathrm{UV}<-17$ galaxies in the full simulation box times the sampled volume. We then calculate the probability of an overdensity in each mock field in the same way as in Sect.~\ref{sec:overdensity}, i.e. for all sampled mock fields and for the subset of mock fields hosting at least one bright $M_\mathrm{UV}<-20.5$ source. 

Qualitatively, we find the same trend as in the observations (Sect.~\ref{sub:bright_overdense}), where mock fields hosting at least one UV-bright galaxy are more likely to be overdense compared to the full sample of fields in all three redshift windows: $7<z<8.5$, $8.5<z<10$, $10<z<11.5$.
Towards higher redshift, where we expect luminous sources to be more clustered as halos become more biased at fixed mass \citep[e.g.,][]{Mo1996}, this effect is enhanced in the simulations. 
In the mock fields, we find that fields hosting a UV-bright source are $1.2-2.1\times$($1.2-1.8\times$) more likely to be overdense at $3\sigma$($2\sigma$) significance compared to a random mock field, with the factor increasing towards higher redshift. 
In the observations, however, we found fields hosting UV-bright galaxies were $2.1-3.2\times$ more likely to be overdense compared to a random field (also finding an increase with redshift, see Sect.~\ref{sub:bright_overdense}).
In the BEACON DR2 sample, bright sources thus appear more clustered than expected from the UV luminosity to halo mass relation from \citet{masonGALAXYUVLUMINOSITY2015}.

To test the impact of enhanced stochasticity, we repeat the procedure of producing 3D galaxy catalogues, this time by adding a scatter of $\sigma_\mathrm{UV}=0.5\,$mag to the UV luminosity to halo mass relation from \citet{masonGALAXYUVLUMINOSITY2015}. Performing the same analysis as before, we find that the relative increase in the fractions is slightly lower: we obtain an increase of $1.1-2.0\times$($1.1-1.8\times$) at $3\sigma$($2\sigma$) significance in this case. Thus, in a simulation with increased scatter in the UV luminosity halo mass relation, bright galaxies are less clustered and thus less likely to be hosted by overdense fields compared to the no UV-scatter scenario. 

\begin{figure}
    \centering
    \includegraphics[width=\columnwidth]{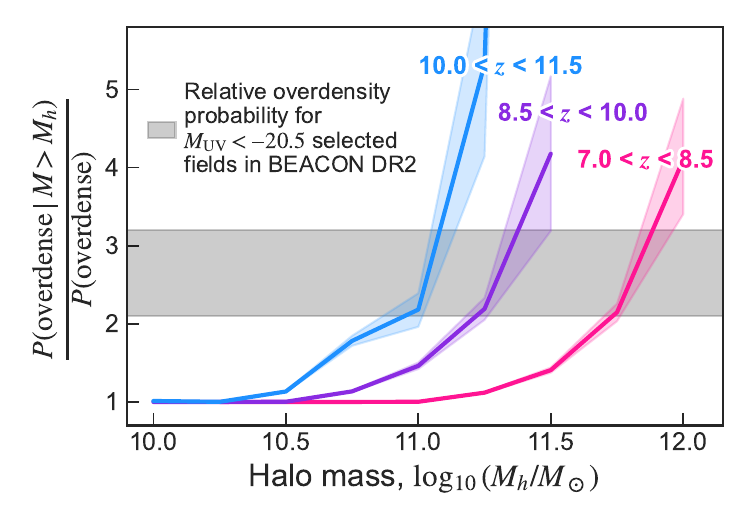}
    \caption{
    Relative overdensity probability, $P(\text{overdense}\,|\,M>M_h) / P(\text{overdense})$, as a function of halo mass threshold, $M_h$, based on 10,000 mock fields from the simulation. This quantifies the excess probability of an overdensity (at the $3\sigma$ level) in a mock field, if it contains a halo more massive than $M_h$.
    Curves show the median result from 500 independent realisations, with shaded regions indicating the 68th percentile range. Results are shown for three redshift windows: $7<z<8.5$ (pink), $8.5<z<10$ (purple), and $10<z<11.5$ (blue). 
    We also mark the range of relative overdensity probabilities, conditional on a field containing an $M_\mathrm{UV}<-20.5$ source, observed in BEACON DR2 as the shaded region.
    }
    \label{fig:overdense_halomass}
\end{figure}

Finally, to assess the characteristic halo mass hosting $M_\mathrm{UV} < -20.5$ galaxies at $z\sim7-11.5$, we analyse the relative overdensity probability, $P(\text{overdense}|M>M_h) / P(\text{overdense})$ in the simulation. This measures how much more likely a field containing a halo more massive than $M_h$ is to be overdense compared to a random mock field. 
For each of the three redshift intervals, $7<z<8.5$, $8.5<z<10$, and $10<z<11.5$, we compute $P(\text{overdense}|M>M_h) / P(\text{overdense})$ by randomly drawing 10,000 mock fields. 
The denominator represents the fraction of all fields that are overdense at the $3\sigma$ level (as defined in Sect.~\ref{sub:bright_overdense}). The numerator is the fraction of fields that host a halo with mass $> M_h$, that are overdense.
We impose a minimum of 3 mock fields satisfying the halo mass criteria for each mass bin to avoid low-number statistics, and rerun the experiment 500 times to estimate uncertainties.

We plot the relative overdensity probability as a function of halo mass from the simulation in Fig.~\ref{fig:overdense_halomass} for the three redshift windows along with the range observed in BEACON for fields hosting $M_\mathrm{UV}<-20.5$ galaxies (see Sect.~\ref{sub:bright_overdense}). The intersection of the curves with the observed BEACON range indicates the halo masses for which the simulations reproduce clustering strengths comparable to those observed, providing an estimate of the halo mass hosting $M_\mathrm{UV}<-20.5$ galaxies.
At $z\sim7.75$, the relative overdensity of $M_\mathrm{UV}<-20.5$ galaxies in BEACON corresponds to halo masses of  $\sim 10^{11.6}\,M_\odot$, whereas at $z\sim10.75$ it drops by almost an order of magnitude, corresponding to halo masses of $\sim 10^{10.8}\,M_\odot$. This suggests that galaxies with similarly bright UV magnitudes are hosted by progressively less massive dark matter halos toward higher redshift. A similar evolutionary trend has recently been reported by \citet{dalmassoAcceleratedEvolutionGalaxy2026} for fainter galaxies, who find that $M_\mathrm{UV}<-17$ galaxies at $z=10.6$ reside in $M_h\sim10^{10.12}M_{\odot}$ halos, compared to $M_h\sim10^{11.45}M_{\odot}$ for $M_\mathrm{UV}<-17$ galaxies at $z=5.5$. The dark matter halo masses found by \citet{dalmassoAcceleratedEvolutionGalaxy2026} are shifted to lower values compared to our results due to their fainter magnitude limit. This decline in halo mass is consistent with a picture of rising accretion rates, and therefore star formation rates, at fixed halo mass at higher redshifts \citep[e.g.,][]{Neistein2006,Correa2015}

In summary, in BEACON DR2 we find that fields hosting $\MUV < -20.5$ galaxies at $z\sim7.0-11.5$ are $2.1-3.2\times$ more likely to be overdense compared to a random field. This is qualitatively consistent with expectations from models where UV luminosity is strongly correlated with halo mass; however, the UV-bright galaxies are more clustered in our observations than in the models we compared to. This may imply UV-bright galaxies are hosted in more massive halos than the \citet{masonGALAXYUVLUMINOSITY2015} model predicts at $z>7$ (with the model predicting $\MUV \sim -20.5$ galaxies reside in $\sim 10^{11}\,M_\odot$ halos at $z\sim10$), and/or a contribution from the one-halo term (i.e. multiple occupation of halos) which is not captured by the simulation \citep[as discussed by][in the context of the angular correlation function at $z>7$]{dalmassoGalaxyClusteringCosmic2024}. Uncertainties in photometric redshifts limit our ability to identify galaxies that are close in 3D space. Spectroscopic follow-up of these overdensities, and more detailed modelling including the one-halo term, will be important to distinguish these effects.
Deeper imaging in a larger number of fields would make it possible to extend this analysis to $z>11.5$, promising new insights into what may be driving the evolution of the LF at the highest redshifts.

\section{Conclusion}
\label{sec:conclusion}

In this paper, we introduce BEACON DR2, the second data release of a JWST Cycle 2 pure-parallel NIRCam imaging survey, consisting of \drNfields\, independent lines of sight, corresponding to a total effective area of 392\,arcmin$^2$ (see also the accompanying paper on galaxy properties, \citealp{zhangBEACONJWSTNIRCam2026}). 
We present selections of galaxy candidates at $z>7$ in these data, and measure the number densities of galaxies at redshifts $z\sim7.5$, $z\sim10$ and $z\sim14$.
Finally, we leverage the large number of independent BEACON fields to identify candidate overdensities and investigate the environments of UV-bright galaxies at these redshifts.
Our main conclusions are as follows:
\begin{enumerate}

    \item We find a total of 164 galaxy candidates at $z>7$. In the three redshift windows defined by our F090W-, F115W-, and F150W-dropout selections, we find, respectively, 150 sources at $z\sim7.5$, 14 sources at $z\sim10$, and no robust $z>13$ candidates.
    
    \item In 11 BEACON pointings overlapping with public JWST spectroscopy, we find no confirmed contaminants in our sample, implying a high purity of our selection. 15 galaxies in our sample are spectroscopically confirmed, with an excellent agreement between photometric and spectroscopic redshifts.
    
    \item We calculate the UV luminosity function in redshift bins corresponding to each dropout selection, finding a decline in number densities with increasing redshift ($z\sim7-14$).

    \item Our UV LF at $z\sim7.5$ is mostly consistent with HST measurements and the few JWST measurements that exist so far, though we find lower number densities in the intermediate magnitude range at $-21\leq M_\mathrm{UV} \leq -19$.
    Our number densities at $z\sim10$ lie below most photometric JWST measurements; however, we find excellent agreement with the spectroscopic measurements by \citet{harikanePureSpectroscopicConstraints2024, harikaneJWSTALMAKeck2025}. This may suggest that selection effects drive the differences between our measured number densities and others in the literature. Our $z>13$ upper limits are consistent with the literature.

    \item We compare our number densities estimated from the three dropout samples with various models. We find that our measurements at redshift $z\sim7.5$ and $z\sim10$ are most compatible with the pre-JWST model by \citet{masonGALAXYUVLUMINOSITY2015}. At $z>13$, our upper limits do not rule out an excess of sources relative to pre-JWST models.

    \item We identify two BEACON fields hosting candidate overdensities at $7<z<8.5$, which are overdense with respect to the Poisson expected number counts at $>3\sigma$ significance: \texttt{beacon\_1420+5252} and \texttt{beacon\_0015-3034}, and one field, \texttt{beacon\_2325-1203} overdense at $2.7\sigma$ significance. 
    The most overdense field, \texttt{beacon\_1420+5252}, overlaps with the EGS field, previously identified as one of the most overdense legacy fields at $z\sim7-9$ \citep[e.g.][]{leonova_prevalence_2022,chenImpactGalaxyOverdensities2026,whitlerDeepJWSTSpectroscopy2026} and likely hosting some of the largest ionised bubbles at these redshifts \citep{Jung2022,tangJWSTNIRSpecSpectroscopy2023a,tangJWSTNIRSpecObservations2024,Napolitano2024,chenImpactGalaxyOverdensities2026}. BEACON adds additional evidence that this field is an important region for understanding the link between large-scale structure and reionisation.

    \item We find significant clustering of UV-bright galaxies in our sample at redshifts $7<z<8.5$ and $8.5<z<10$. Fields hosting at least one $M_\mathrm{UV} < -20.5$ galaxy are approximately three times as likely to be overdense compared to our full sample of fields. This is qualitatively consistent with expectations from models where UV luminosity is strongly correlated with halo mass; however, we find that the UV-bright galaxies in our sample are more strongly clustered than expected by the models. 
    Comparison with mock halo catalogues based on semi-numerical simulations suggests the $M_\mathrm{UV} < -20.5$ sources inhabit halos of $M_h \sim 10^{11.6} M_\odot$ at $z \sim 8$, decreasing to $M_h \sim 10^{10.8} M_\odot$ by $z \sim 11$, consistent with a picture of increasing star formation rates at higher redshift, and/or may point to multiple halo occupation.
\end{enumerate}

In the future, deep spectroscopic coverage of $z>7$ candidates with JWST will be crucial for refining photometric selections of candidates at high redshift and will facilitate more dedicated comparisons with HST selections in order to more accurately measure the high redshift UV LF.
In addition, ongoing efforts with both pure-parallel and prime programs to obtain JWST NIRCam imaging over hundreds of independent fields will build a legacy of deep fields,  enabling more quantitative clustering analyses to shed light on the galaxy-halo connection in the first billion years.

\begin{acknowledgements}
      KCK, CAM and VG acknowledge support from the Carlsberg Foundation under grant CF22-1322. CAM also acknowledges support by the European Union ERC grant RISES (101163035) and VILLUM FONDEN (37459). Views and opinions expressed are those of the author(s) only and do not necessarily reflect those of the European Union or the European Research Council. Neither the European Union nor the granting authority can be held responsible for them.
      The Cosmic Dawn Center (DAWN) is funded by the Danish National Research Foundation under grant DNRF140. MS is partially supported by NASA grant 80NSSC22K1294. MB acknowledges support from the ERC Grant FIRSTLIGHT, and Slovenian national research agency ARIS through grants N1-0238 and P1-0188. AJB acknowledges funding from the “FirstGalaxies” Advanced Grant from the European Research Council (ERC) under the European Union’s Horizon 2020 research and innovation program (Grant agreement No. 789056). 
      HA acknowledges support from CNES, focused on the JWST mission, and the Programme National Cosmology and Galaxies (PNCG) of CNRS/INSU with INP and IN2P3, co-funded by CEA and CNES and support by the French National Research Agency (ANR) under grant ANR-21-CE31-0838. Support for JWST program 3990 was provided by NASA through the Space Telescope Science Institute, which is operated by the Association of Universities for Research in Astronomy, Inc., under NASA contract NAS 5-03127. All of the data presented in this paper were obtained from the Mikulski Archive for Space Telescopes (MAST) at the Space Telescope Science Institute. BEACON observations analysed can be accessed via DOI:\href{https://doi.org/10.17909/5x1p-dp20}{10.17909/5x1p-dp20}. All our data products from BEACON DR2 are available at MAST as a High Level Science Product via \href{https://doi.org/10.17909/f2rf-hc66}{10.17909/f2rf-hc66}. 
      The Tycho supercomputer hosted at the SCIENCE HPC center at the University of Copenhagen and DeiC National HPC (via project Deic-KU-L5-2025-016-001) were used for supporting this work. Some of the data products presented herein were retrieved from the Dawn JWST Archive (DJA). DJA is an initiative of the Cosmic Dawn Center (DAWN), which is funded by the Danish National Research Foundation under grant DNRF140.
      This work made use of the following open-source software: \texttt{Matplotlib} \citep{hunterMatplotlib2DGraphics2007}, \texttt{Numpy} \citep{harrisArrayProgrammingNumPy2020}, \texttt{SciPy} \citep{virtanenSciPy10Fundamental2020}, \texttt{Astropy} \citep{theastropycollaborationAstropyProjectSustaining2022}, \texttt{emcee} \cite{foreman-mackeyEmceeMCMCHammer2013}, \texttt{pandas} \citep{thepandasdevelopmentteamPandasdevPandasPandas2024}, \texttt{Corner} \citep{foreman-mackeyCornerpyScatterplotMatrices2016}, and \texttt{Numba} \citep{lamNumbaLLVMbasedPython2015}.
\end{acknowledgements}

\bibliographystyle{aasjournalv7}
\bibliography{references,library_cm}

\begin{appendix}
\section{Field and candidate tables}

We provide an overview of the 36 BEACON DR2 fields used in this study in Tab.~\ref{app:field_tables}, specifying the field ID, the position, and the 5$\sigma$ limiting magnitudes, measured within a $0.\!''16$ aperture radius, for all filters available in a given field.

We list our F090W-dropout candidates in Tab.~\ref{tab:f090w_dropouts_1}--\ref{tab:f090w_dropouts_5}, and our F115W-dropout candidates in Tab.~\ref{tab:f115w_dropouts_1}.
For each candidate, we specify the field and object ID, the position, the absolute UV magnitude (corrected for lensing in the field \texttt{beacon\_0014-3025}, which overlaps with the Abell-2744 field), and the photometric redshift from \texttt{EAZY}. If available, the spectroscopic redshift from the DJA is listed along with information about the relevant spectroscopic program.

\label{app:field_tables}
\AddToHook{env/tabular/begin}[sizehook]{\footnotesize}
\vspace{0.5cm}
\begin{deluxetable*}{lcccccccccccccccccc}
\tablecaption{5$\sigma$ limiting magnitudes for point sources in each of the available filters for each BEACON DR2 field\label{tab:field_depths}}

\tabletypesize{\scriptsize}
\setlength{\tabcolsep}{3pt}
\renewcommand{\arraystretch}{1.25}

\tablehead{
\colhead{Field ID} &
\colhead{RA} &
\colhead{DEC} &
\colhead{F070W} &
\colhead{F090W} &
\colhead{F115W} &
\colhead{F140M} &
\colhead{F150W} &
\colhead{F182M} &
\colhead{F200W} &
\colhead{F277W} &
\colhead{F335M} &
\colhead{F356W} &
\colhead{F360M} &
\colhead{F410M} &
\colhead{F430M} &
\colhead{F444W} &
\colhead{F460M} &
\colhead{F480M}
}

\startdata
1420+5253 & 214.92971 & 52.89569 & 28.0 & 28.8 & 28.7 & 28.0 & 28.7 & 28.3 & 28.8 & 29.0 & 28.1 & 29.0 & 28.1 & 28.3 & 27.7 & 28.6 & 27.0 & 27.5 \\
0447-2637$^\dagger$ & 71.67564 & -26.60023 & - & 27.6 & 27.4 & 27.1 & 27.9 & 27.5 & 28.1 & 28.5 & - & 28.5 & - & 27.8 & 27.0 & 28.1 & - & 26.8 \\
1420+5252$^\dagger$ & 215.07468 & 52.86999 & - & 28.0 & 28.1 & 27.5 & 28.3 & 27.8 & 28.5 & 28.9 & - & 28.9 & - & 28.2 & 27.5 & 28.6 & - & 27.3 \\
0015-3034 & 3.67222 & -30.55468 & - & 28.1 & 28.2 & 27.7 & 28.4 & - & 28.6 & 28.8 & - & 28.9 & - & 28.1 & 27.3 & 28.6 & - & - \\
1230+2702 & 187.47251 & 27.02898 & - & 26.2 & 25.9 & - & 26.2 & 25.6 & 26.2 & 27.3 & - & 27.4 & - & 26.9 & - & 27.3 & - & 25.6 \\
0332-2749$^\dagger$ & 53.03300 & -27.81392 & - & 29.4 & 29.4 & - & 28.0 & - & 29.7 & 29.4 & - & 28.5 & - & 29.1 & - & 29.3 & - & - \\
0959+0200$^\dagger$ & 149.84149 & 2.01664 & - & 28.3 & 28.3 & - & 28.5 & - & 28.6 & 28.9 & - & 28.9 & - & 28.2 & - & 28.5 & - & - \\
1329+4707 & 202.22921 & 47.12018 & - & 26.1 & 26.2 & - & 26.7 & - & 26.8 & 27.9 & - & 27.9 & - & 27.1 & - & 27.7 & - & - \\
1329+4709 & 202.28636 & 47.14673 & - & 26.0 & 26.1 & - & 26.6 & - & 26.7 & 27.9 & - & 27.9 & - & 27.0 & - & 27.6 & - & - \\
2304-6250 & 345.96262 & -62.83896 & - & 26.8 & 26.4 & - & 27.9 & - & 28.0 & 28.6 & - & 28.8 & - & 27.5 & - & 27.9 & - & - \\
2325-1203 & 351.28991 & -12.04149 & - & 27.9 & 27.9 & - & 28.2 & - & 28.3 & 28.6 & - & 28.7 & - & 27.9 & - & 28.4 & - & - \\
0014-3025$^\dagger$ & 3.59520 & -30.41974 & - & 28.9 & 28.8 & - & - & - & 29.1 & 29.1 & - & 29.3 & - & - & - & 29.0 & - & - \\
0055-3749 & 13.84625 & -37.80868 & - & 27.6 & 27.6 & - & 27.9 & - & - & 28.5 & - & 28.6 & - & - & - & 28.2 & - & - \\
0217-0504$^\dagger$ & 34.30829 & -5.07792 & - & 29.2 & 29.0 & - & 29.2 & - & - & 29.5 & - & 29.6 & - & - & - & 29.3 & - & - \\
0217-0508$^\dagger$ & 34.29253 & -5.12272 & - & 29.4 & 29.0 & - & 29.2 & - & - & 29.2 & - & 29.2 & - & - & - & 28.9 & - & - \\
0217-0509$^\dagger$ & 34.21885 & -5.13408 & - & 29.3 & 29.0 & - & 29.2 & - & - & 29.1 & - & 29.2 & - & - & - & 28.8 & - & - \\
0227-5319 & 36.83709 & -53.32894 & - & 28.1 & 28.2 & - & 28.4 & - & - & 28.9 & - & 28.9 & - & - & - & 28.6 & - & - \\
0240-0253$^\dagger$ & 40.09595 & -2.87019 & - & 26.9 & 26.9 & - & 27.2 & - & - & 28.0 & - & 28.1 & - & - & - & 27.5 & - & - \\
0303+0060 & 45.73565 & -1.00921 & - & 28.2 & 28.3 & - & 28.5 & - & - & 29.0 & - & 29.1 & - & - & - & 28.6 & - & - \\
0332-2745$^\dagger$ & 53.03766 & -27.75151 & - & 28.9 & 29.0 & - & 29.1 & - & 29.3 & 29.6 & - & 29.5 & - & - & - & 29.2 & - & - \\
0338-0454 & 54.45726 & -4.91147 & - & 27.9 & 28.0 & - & 28.2 & - & - & 28.7 & - & 28.7 & - & - & - & 28.2 & - & - \\
0442-4613$^\dagger$ & 70.42773 & -46.21713 & - & 26.6 & 26.1 & - & 26.4 & - & - & 27.5 & - & 27.5 & - & - & - & 27.7 & - & - \\
0502-4338$^\dagger$ & 75.45100 & -43.63230 & - & - & 27.5 & - & 27.8 & - & 28.0 & 28.5 & - & 28.5 & - & - & - & 28.2 & - & - \\
0843+0324$^\dagger$ & 130.65348 & 3.41877 & - & - & 27.6 & - & 27.8 & - & 27.9 & 28.3 & - & 28.3 & - & - & - & 27.8 & - & - \\
0860+3857$^\dagger$ & 134.98519 & 38.96236 & - & 27.9 & 27.5 & - & 27.7 & - & - & 28.5 & - & 28.5 & - & - & - & 28.2 & - & - \\
1010+2701$^\dagger$ & 152.44541 & 27.03744 & - & 28.1 & 28.1 & - & 28.4 & - & - & 28.7 & - & 28.7 & - & - & - & 28.3 & - & - \\
1114+3235 & 168.53729 & 32.60011 & - & 27.1 & 27.1 & - & 27.4 & - & - & 28.0 & - & 28.1 & - & - & - & 27.7 & - & - \\
1138+5748$^\dagger$ & 174.41601 & 57.80085 & - & 27.4 & 27.3 & - & 27.6 & - & - & 28.5 & - & 28.4 & - & - & - & 28.1 & - & - \\
1227+2157 & 186.76729 & 21.94950 & - & - & 28.4 & - & 28.5 & - & 28.7 & 28.8 & - & 29.0 & - & - & - & 28.5 & - & - \\
1526+3559 & 231.59796 & 35.98129 & - & 26.7 & 26.8 & - & 27.0 & - & - & 27.9 & - & 28.0 & - & - & - & 27.5 & - & - \\
1526+3560 & 231.58035 & 35.99999 & - & 27.1 & 27.1 & - & 27.5 & - & - & 28.2 & - & 28.3 & - & - & - & 27.9 & - & - \\
2059-4247$^\dagger$ & 314.63703 & -42.78990 & - & - & 27.0 & - & 27.3 & - & 27.4 & 28.0 & - & 28.1 & - & - & - & 27.6 & - & - \\
2252-1744 & 342.96081 & -17.73690 & - & 28.2 & 27.7 & - & 28.0 & - & - & 28.5 & - & 28.7 & - & - & - & 28.7 & - & - \\
2316-5910$^\dagger$ & 349.11637 & -59.15965 & - & - & 27.0 & - & 27.3 & - & 27.5 & 28.1 & - & 28.2 & - & - & - & 27.8 & - & - \\
2325-1216$^\dagger$ & 351.35492 & -12.26333 & - & - & 26.5 & - & 26.8 & - & 26.8 & 27.7 & - & 27.8 & - & - & - & 27.6 & - & - \\
2346+1256 & 356.60030 & 12.92787 & - & 27.6 & 27.8 & - & 28.0 & - & - & 28.5 & - & 28.6 & - & - & - & 28.0 & - & - \\
\enddata

\tablenotetext{}{$^\dagger$ indicates BEACON DR1 fields}

\end{deluxetable*}

\RemoveFromHook{env/tabular/begin}[sizehook]

\begin{deluxetable*}{llcccccc}
\renewcommand{\arraystretch}{1.5} 
\caption{F090W-dropouts}

\tablehead{
\colhead{Field ID} &
\colhead{Object ID} &
\colhead{R.A.} &
\colhead{Decl.} &
\colhead{$M_{\mathrm{UV}}$} &
\colhead{$z_{\mathrm{phot}}$} &
\colhead{$z_{\mathrm{spec}}$} &
\colhead{Spectroscopic Program}
}

\startdata
0217-0509 & 1447 & 34.225235 & -5.182756 & $-18.66^{+0.21}_{-0.20}$ & $6.82^{+0.23}_{-0.16}$ & $-$ & $-$ \\
0217-0508 & 15564 & 34.285568 & -5.103937 & $-18.19^{+0.47}_{-0.35}$ & $6.89^{+0.74}_{-0.61}$ & $-$ & $-$ \\
2252-1744 & 567 & 342.954895 & -17.768801 & $-20.59^{+0.27}_{-0.15}$ & $6.93^{+0.12}_{-0.20}$ & $-$ & $-$ \\
1420+5253 & 3083 & 214.893906 & 52.874592 & $-19.77^{+0.12}_{-0.13}$ & $6.99^{+-0.03}_{-0.11}$ & $7.031$ & EGS, RUBIES (GO-4233, PI: A. de Graaff) [1][2] \\
1420+5252 & 3062 & 215.040863 & 52.863846 & $-19.30^{+0.31}_{-0.30}$ & $7.00^{+0.13}_{-0.11}$ & $-$ & $-$ \\
1420+5252 & 7386 & 215.037430 & 52.877472 & $-18.73^{+0.28}_{-0.24}$ & $7.02^{+0.02}_{-0.14}$ & $-$ & $-$ \\
1420+5252 & 10296 & 215.125000 & 52.874062 & $-20.79^{+0.10}_{-0.11}$ & $7.05^{+0.16}_{-0.16}$ & $-$ & $-$ \\
0015-3034 & 3875 & 3.695413 & -30.570082 & $-19.83^{+0.12}_{-0.12}$ & $7.05^{+-0.00}_{-0.16}$ & $-$ & $-$ \\
1420+5253 & 3711 & 214.892609 & 52.880627 & $-19.48^{+0.22}_{-0.27}$ & $7.06^{+0.07}_{-0.17}$ & $6.993$ & EGS, RUBIES (GO-4233, PI: A. de Graaff) [1][2] \\
2325-1203 & 1860 & 351.302551 & -12.079363 & $-20.35^{+0.16}_{-0.14}$ & $7.06^{+0.32}_{-0.09}$ & $-$ & $-$ \\
2304-6250 & 22812 & 345.940277 & -62.829922 & $-20.22^{+0.20}_{-0.21}$ & $7.09^{+0.29}_{-0.12}$ & $-$ & $-$ \\
0217-0504 & 18177 & 34.323090 & -5.081106 & $-19.29^{+0.22}_{-0.18}$ & $7.09^{+0.81}_{-0.12}$ & $-$ & $-$ \\
0015-3034 & 7054 & 3.682551 & -30.541414 & $-18.57^{+0.32}_{-0.28}$ & $7.09^{+0.12}_{-0.12}$ & $-$ & $-$ \\
0015-3034 & 2121 & 3.705384 & -30.578817 & $-19.11^{+0.29}_{-0.27}$ & $7.09^{+0.12}_{-0.12}$ & $-$ & $-$ \\
0015-3034 & 12964 & 3.645315 & -30.543470 & $-18.81^{+0.50}_{-0.30}$ & $7.11^{+0.10}_{-0.22}$ & $-$ & $-$ \\
0860+3857 & 14167 & 134.982468 & 38.985676 & $-20.00^{+0.24}_{-0.20}$ & $7.11^{+0.69}_{-0.07}$ & $-$ & $-$ \\
0217-0504 & 6307 & 34.273205 & -5.037470 & $-18.69^{+0.18}_{-0.18}$ & $7.13^{+0.17}_{-0.32}$ & $-$ & $-$ \\
1420+5252 & 7298 & 215.038239 & 52.877132 & $-19.37^{+0.30}_{-0.29}$ & $7.13^{+0.00}_{-0.24}$ & $7.176$ & EGS, RUBIES (GO-4233, PI: A. de Graaff) [1][2] \\
1138+5748 & 6058 & 174.370514 & 57.791782 & $-20.13^{+0.26}_{-0.19}$ & $7.14^{+0.16}_{-0.17}$ & $-$ & $-$ \\
0959+0200 & 14013 & 149.833344 & 2.036179 & $-19.35^{+0.20}_{-0.18}$ & $7.14^{+0.23}_{-0.25}$ & $-$ & $-$ \\
0332-2745 & 7700 & 53.071823 & -27.782648 & $-19.31^{+0.12}_{-0.11}$ & $7.16^{+0.39}_{-0.11}$ & $-$ & $-$ \\
1420+5252 & 12575 & 215.025375 & 52.891273 & $-19.59^{+0.16}_{-0.19}$ & $7.16^{+-0.03}_{-0.19}$ & $-$ & $-$ \\
1420+5252 & 12023 & 215.039520 & 52.891056 & $-18.83^{+0.29}_{-0.30}$ & $7.17^{+-0.04}_{-0.28}$ & $-$ & $-$ \\
0015-3034 & 3734 & 3.699224 & -30.568316 & $-19.50^{+0.19}_{-0.21}$ & $7.17^{+0.04}_{-0.21}$ & $-$ & $-$ \\
1420+5252 & 9579 & 215.037079 & 52.892605 & $-19.65^{+0.23}_{-0.26}$ & $7.18^{+0.03}_{-0.21}$ & $7.196$ & EGS, RUBIES (GO-4233, PI: A. de Graaff) [1][2] \\
0015-3034 & 3690 & 3.699117 & -30.568407 & $-19.89^{+0.17}_{-0.14}$ & $7.19^{+0.02}_{-0.22}$ & $-$ & $-$ \\
1420+5252 & 11933 & 215.035583 & 52.892235 & $-21.71^{+0.07}_{-0.06}$ & $7.20^{+0.01}_{-0.15}$ & $-$ & $-$ \\
0217-0504 & 6678 & 34.284660 & -5.053731 & $-18.06^{+0.48}_{-0.28}$ & $7.22^{+0.41}_{-0.41}$ & $-$ & $-$ \\
0217-0504 & 9016 & 34.283463 & -5.046202 & $-19.07^{+0.19}_{-0.19}$ & $7.24^{+2.62}_{-0.03}$ & $-$ & $-$ \\
0217-0509 & 2696 & 34.237087 & -5.167516 & $-19.65^{+0.27}_{-0.25}$ & $7.24^{+2.73}_{--0.13}$ & $-$ & $-$ \\
2346+1256 & 15536 & 356.575745 & 12.970267 & $-19.59^{+0.57}_{-0.30}$ & $7.26^{+0.46}_{-0.83}$ & $-$ & $-$ \\
1010+2701 & 7624 & 152.441269 & 27.018648 & $-19.47^{+0.51}_{-0.39}$ & $7.26^{+0.81}_{-0.30}$ & $-$ & $-$ \\
0217-0509 & 336 & 34.223042 & -5.191769 & $-20.96^{+0.06}_{-0.04}$ & $7.27^{+0.02}_{-0.14}$ & $-$ & $-$ \\
0217-0508 & 8517 & 34.283566 & -5.124747 & $-18.05^{+0.27}_{-0.25}$ & $7.27^{+0.10}_{-0.22}$ & $-$ & $-$ \\
1420+5252 & 1569 & 214.994904 & 52.866634 & $-19.22^{+0.66}_{-0.31}$ & $7.28^{+-0.07}_{-0.31}$ & $-$ & $-$ \\
\enddata

\tablenotetext{}{[1] \citet{degraaffRUBIESCompleteCensus2025}; [2] \citet{chenImpactGalaxyOverdensities2026}}
\label{tab:f090w_dropouts_1}
\end{deluxetable*}


\begin{deluxetable*}{llcccccc}
\renewcommand{\arraystretch}{1.5} 
\caption{F090W-dropouts (continued)}

\tablehead{
\colhead{Field ID} &
\colhead{Object ID} &
\colhead{R.A.} &
\colhead{Decl.} &
\colhead{$M_{\mathrm{UV}}$} &
\colhead{$z_{\mathrm{phot}}$} &
\colhead{$z_{\mathrm{spec}}$} &
\colhead{Spectroscopic Program}
}

\startdata
0217-0509 & 435 & 34.222919 & -5.191683 & $-18.66^{+0.20}_{-0.23}$ & $7.28^{+0.10}_{-0.15}$ & $-$ & $-$ \\
0332-2745 & 15094 & 53.046597 & -27.738073 & $-19.80^{+0.08}_{-0.08}$ & $7.29^{+0.17}_{-0.16}$ & $-$ & $-$ \\
0217-0509 & 12575 & 34.207989 & -5.121376 & $-18.80^{+0.20}_{-0.18}$ & $7.29^{+0.34}_{-0.24}$ & $-$ & $-$ \\
1138+5748 & 2321 & 174.362198 & 57.776512 & $-20.29^{+0.42}_{-0.44}$ & $7.29^{+3.02}_{-0.08}$ & $-$ & $-$ \\
0217-0508 & 12114 & 34.277187 & -5.113205 & $-18.72^{+0.23}_{-0.29}$ & $7.30^{+1.33}_{-0.17}$ & $-$ & $-$ \\
0332-2749 & 5025 & 53.086330 & -27.823927 & $-19.68^{+0.21}_{-0.16}$ & $7.32^{+0.23}_{-0.19}$ & $7.605$ & GOODS-S, GO-2198 (PI: L. Barrufet) [1] \\
1010+2701 & 3159 & 152.407913 & 27.005966 & $-20.22^{+0.18}_{-0.14}$ & $7.33^{+0.13}_{-0.28}$ & $-$ & $-$ \\
1114+3235 & 15133 & 168.548447 & 32.625988 & $-20.42^{+0.22}_{-0.22}$ & $7.34^{+1.49}_{-0.21}$ & $-$ & $-$ \\
0303+0060 & 3520 & 45.743046 & -1.021974 & $-19.46^{+0.22}_{-0.17}$ & $7.34^{+0.38}_{-0.29}$ & $-$ & $-$ \\
1526+3560 & 20223 & 231.601318 & 36.017792 & $-20.27^{+0.41}_{-0.28}$ & $7.34^{+1.49}_{-0.21}$ & $-$ & $-$ \\
2252-1744 & 8478 & 342.954559 & -17.715113 & $-20.80^{+0.12}_{-0.17}$ & $7.34^{+0.55}_{-0.21}$ & $-$ & $-$ \\
0055-3749 & 9055 & 13.864957 & -37.833458 & $-19.88^{+0.43}_{-0.24}$ & $7.34^{+0.38}_{-0.45}$ & $-$ & $-$ \\
0014-3025 & 10125 & 3.625879 & -30.439075 & $-20.06^{+0.12}_{-0.17}$ & $7.34^{+-0.05}_{-0.13}$ & $-$ & $-$ \\
2325-1203 & 1657 & 351.317657 & -12.072651 & $-19.50^{+0.21}_{-0.21}$ & $7.34^{+-0.05}_{-0.30}$ & $-$ & $-$ \\
0227-5319 & 17480 & 36.867207 & -53.278366 & $-19.13^{+0.34}_{-0.26}$ & $7.34^{+0.91}_{-0.14}$ & $-$ & $-$ \\
0014-3025 & 11608 & 3.612495 & -30.427191 & $-19.77^{+0.23}_{-0.23}$ & $7.35^{+0.91}_{-0.14}$ & $-$ & $-$ \\
2325-1203 & 1602 & 351.311798 & -12.076117 & $-20.95^{+0.09}_{-0.07}$ & $7.35^{+-0.06}_{-0.30}$ & $-$ & $-$ \\
0217-0504 & 3421 & 34.270123 & -5.040110 & $-19.47^{+0.14}_{-0.12}$ & $7.35^{+0.63}_{-0.30}$ & $-$ & $-$ \\
1010+2701 & 16958 & 152.464661 & 27.046890 & $-19.26^{+1.28}_{-0.30}$ & $7.35^{+1.09}_{-4.23}$ & $-$ & $-$ \\
0332-2745 & 8900 & 53.053425 & -27.765003 & $-19.22^{+0.16}_{-0.14}$ & $7.35^{+0.02}_{-0.22}$ & $-$ & $-$ \\
0303+0060 & 2610 & 45.729721 & -1.029094 & $-21.46^{+0.04}_{-0.04}$ & $7.35^{+-0.06}_{-0.22}$ & $-$ & $-$ \\
2325-1203 & 740 & 351.295288 & -12.091648 & $-18.96^{+0.34}_{-0.36}$ & $7.35^{+0.02}_{-0.38}$ & $-$ & $-$ \\
0217-0509 & 1793 & 34.224037 & -5.180541 & $-19.13^{+0.15}_{-0.13}$ & $7.36^{+0.81}_{-0.06}$ & $-$ & $-$ \\
0217-0508 & 8334 & 34.286720 & -5.123929 & $-17.63^{+0.52}_{-0.27}$ & $7.36^{+0.81}_{-0.39}$ & $-$ & $-$ \\
0332-2745 & 8698 & 53.053013 & -27.765186 & $-19.14^{+0.18}_{-0.16}$ & $7.36^{+0.44}_{-0.23}$ & $-$ & $-$ \\
0014-3025 & 14457 & 3.608568 & -30.418518 & $-21.84^{+0.05}_{-0.06}$ & $7.36^{+0.27}_{-0.15}$ & $7.286$ & Abell-2744, GLASS (GO 1324, PI: T. Treu) [2] \\
2252-1744 & 12537 & 342.963623 & -17.686695 & $-19.47^{+0.32}_{-0.24}$ & $7.36^{+0.44}_{-0.32}$ & $-$ & $-$ \\
0217-0504 & 6193 & 34.273129 & -5.037531 & $-18.76^{+0.17}_{-0.18}$ & $7.38^{+0.52}_{-0.25}$ & $-$ & $-$ \\
2325-1203 & 7977 & 351.284821 & -12.030617 & $-19.87^{+0.16}_{-0.16}$ & $7.38^{+-0.00}_{-0.33}$ & $-$ & $-$ \\
0014-3025 & 2599 & 3.553823 & -30.410107 & $-18.95^{+0.33}_{-0.25}$ & $7.39^{+0.69}_{-0.26}$ & $-$ & $-$ \\
2325-1203 & 11012 & 351.274292 & -12.017154 & $-19.68^{+0.23}_{-0.15}$ & $7.39^{+-0.01}_{-0.42}$ & $-$ & $-$ \\
0959+0200 & 1159 & 149.840607 & 1.964149 & $-19.15^{+0.23}_{-0.24}$ & $7.40^{+-0.02}_{-0.35}$ & $-$ & $-$ \\
0303+0060 & 8459 & 45.747028 & -0.986697 & $-20.28^{+0.10}_{-0.10}$ & $7.42^{+0.48}_{-0.12}$ & $-$ & $-$ \\
0015-3034 & 4322 & 3.685795 & -30.573059 & $-19.02^{+0.34}_{-0.28}$ & $7.42^{+0.04}_{-0.12}$ & $-$ & $-$ \\
0332-2745 & 13954 & 53.086193 & -27.773207 & $-18.52^{+0.31}_{-0.19}$ & $7.43^{+0.29}_{-0.30}$ & $-$ & $-$ \\
\enddata

\tablenotetext{}{[1] \citet{barrufetQuiescentDustyUnveiling2025}; [2] \citet{masciaGLASSJWSTEarlyRelease2024}}
\label{tab:f090w_dropouts_2}
\end{deluxetable*}


\begin{deluxetable*}{llcccccc}
\renewcommand{\arraystretch}{1.5} 
\caption{F090W-dropouts (continued)}

\tablehead{
\colhead{Field ID} &
\colhead{Object ID} &
\colhead{R.A.} &
\colhead{Decl.} &
\colhead{$M_{\mathrm{UV}}$} &
\colhead{$z_{\mathrm{phot}}$} &
\colhead{$z_{\mathrm{spec}}$} &
\colhead{Spectroscopic Program}
}

\startdata
1420+5252 & 3630 & 215.084747 & 52.858551 & $-19.49^{+0.20}_{-0.16}$ & $7.43^{+0.02}_{-0.14}$ & $-$ & $-$ \\
0303+0060 & 2693 & 45.729534 & -1.029183 & $-18.95^{+0.53}_{-0.34}$ & $7.46^{+0.99}_{-0.49}$ & $-$ & $-$ \\
0217-0509 & 1974 & 34.221172 & -5.180426 & $-19.79^{+0.08}_{-0.07}$ & $7.46^{+0.09}_{-0.25}$ & $-$ & $-$ \\
2325-1203 & 1556 & 351.311035 & -12.076932 & $-19.81^{+0.22}_{-0.22}$ & $7.46^{+-0.08}_{-0.41}$ & $-$ & $-$ \\
0959+0200 & 92 & 149.819672 & 1.965411 & $-19.89^{+0.17}_{-0.14}$ & $7.46^{+-0.09}_{-0.57}$ & $-$ & $-$ \\
0303+0060 & 9711 & 45.724251 & -0.983340 & $-19.03^{+0.28}_{-0.24}$ & $7.48^{+-0.02}_{-0.43}$ & $-$ & $-$ \\
0332-2745 & 9353 & 53.022465 & -27.739815 & $-18.27^{+0.26}_{-0.21}$ & $7.50^{+0.40}_{-0.45}$ & $-$ & $-$ \\
0959+0200 & 14134 & 149.860275 & 2.027343 & $-19.55^{+0.48}_{-0.29}$ & $7.50^{+-0.04}_{-0.62}$ & $-$ & $-$ \\
0332-2749 & 24726 & 53.001209 & -27.802843 & $-20.07^{+0.14}_{-0.19}$ & $7.51^{+-0.05}_{-0.30}$ & $-$ & $-$ \\
0217-0508 & 1040 & 34.320229 & -5.164967 & $-18.99^{+0.31}_{-0.20}$ & $7.51^{+0.38}_{-0.46}$ & $-$ & $-$ \\
0217-0508 & 3329 & 34.318546 & -5.148247 & $-18.79^{+0.24}_{-0.17}$ & $7.53^{+0.45}_{-0.40}$ & $-$ & $-$ \\
0217-0509 & 1420 & 34.225983 & -5.182612 & $-18.49^{+0.41}_{-0.25}$ & $7.54^{+0.53}_{-0.50}$ & $-$ & $-$ \\
0217-0509 & 85 & 34.234711 & -5.188540 & $-19.10^{+0.34}_{-0.24}$ & $7.58^{+0.31}_{-0.70}$ & $-$ & $-$ \\
0217-0508 & 8355 & 34.290524 & -5.122096 & $-18.33^{+0.25}_{-0.23}$ & $7.59^{+0.30}_{-0.62}$ & $-$ & $-$ \\
0217-0509 & 2346 & 34.244049 & -5.167047 & $-17.96^{+0.42}_{-0.34}$ & $7.62^{+0.54}_{-0.89}$ & $-$ & $-$ \\
2252-1744 & 8464 & 342.941772 & -17.719185 & $-20.20^{+0.31}_{-0.19}$ & $7.63^{+0.44}_{-0.34}$ & $-$ & $-$ \\
1420+5252 & 12428 & 215.084641 & 52.882202 & $-19.27^{+0.20}_{-0.20}$ & $7.68^{+-0.05}_{-0.13}$ & $-$ & $-$ \\
1420+5253 & 8384 & 214.952133 & 52.899982 & $-20.36^{+0.10}_{-0.06}$ & $7.70^{+-0.07}_{-0.07}$ & $-$ & $-$ \\
0332-2745 & 18704 & 53.035835 & -27.718058 & $-18.72^{+0.35}_{-0.22}$ & $7.70^{+0.37}_{-0.82}$ & $-$ & $-$ \\
0332-2745 & 15255 & 53.043163 & -27.734766 & $-18.09^{+0.41}_{-0.30}$ & $7.71^{+0.45}_{-0.75}$ & $-$ & $-$ \\
0015-3034 & 5857 & 3.658228 & -30.566301 & $-20.12^{+0.07}_{-0.07}$ & $7.74^{+-0.02}_{-0.11}$ & $-$ & $-$ \\
0217-0508 & 14393 & 34.269173 & -5.105772 & $-18.92^{+0.15}_{-0.13}$ & $7.80^{+0.19}_{-0.67}$ & $-$ & $-$ \\
0217-0504 & 15163 & 34.292065 & -5.043071 & $-19.35^{+0.13}_{-0.10}$ & $7.81^{+0.18}_{-0.51}$ & $-$ & $-$ \\
0959+0200 & 8926 & 149.829056 & 2.021058 & $-19.19^{+0.20}_{-0.15}$ & $7.84^{+0.14}_{-0.12}$ & $-$ & $-$ \\
0217-0509 & 9363 & 34.219463 & -5.127748 & $-18.97^{+0.11}_{-0.10}$ & $7.85^{+0.13}_{-0.73}$ & $6.990$ & UDS, RUBIES (GO-4233, PI: A. de Graaff) [1] \\
0217-0508 & 3572 & 34.296047 & -5.156882 & $-19.10^{+0.21}_{-0.14}$ & $7.87^{+0.21}_{-0.74}$ & $-$ & $-$ \\
0332-2749 & 12382 & 53.087475 & -27.814884 & $-19.79^{+0.16}_{-0.16}$ & $7.93^{+0.05}_{-0.21}$ & $-$ & $-$ \\
0217-0509 & 6584 & 34.210968 & -5.142348 & $-18.04^{+0.43}_{-0.34}$ & $7.94^{+0.50}_{-0.73}$ & $-$ & $-$ \\
0332-2749 & 18718 & 53.058949 & -27.808054 & $-20.04^{+0.12}_{-0.11}$ & $7.95^{+0.13}_{-0.23}$ & $-$ & $-$ \\
0217-0508 & 1995 & 34.301559 & -5.166206 & $-19.52^{+0.12}_{-0.13}$ & $7.95^{+0.12}_{-0.74}$ & $6.996$ & UDS, RUBIES (GO-4233, PI: A. de Graaff) [1] \\
0014-3025 & 8648 & 3.613551 & -30.434706 & $-20.07^{+0.17}_{-0.14}$ & $7.97^{+0.11}_{-0.67}$ & $-$ & $-$ \\
0015-3034 & 5505 & 3.664271 & -30.563770 & $-19.11^{+0.25}_{-0.21}$ & $7.97^{+0.38}_{-0.17}$ & $-$ & $-$ \\
0055-3749 & 5433 & 13.891001 & -37.856560 & $-20.30^{+0.20}_{-0.15}$ & $8.02^{+0.33}_{-0.81}$ & $-$ & $-$ \\
0332-2745 & 16560 & 53.084354 & -27.761089 & $-18.45^{+0.21}_{-0.20}$ & $8.03^{+0.32}_{-0.82}$ & $-$ & $-$ \\
0217-0509 & 9252 & 34.209660 & -5.132755 & $-17.79^{+0.43}_{-0.35}$ & $8.07^{+0.47}_{-1.10}$ & $-$ & $-$ \\
\enddata
\tablenotetext{}{[1] \citet{degraaffRUBIESCompleteCensus2025}}
\label{tab:f090w_dropouts_3}
\end{deluxetable*}


\begin{deluxetable*}{llcccccc}
\renewcommand{\arraystretch}{1.5} 
\caption{F090W-dropouts (continued)}

\tablehead{
\colhead{Field ID} &
\colhead{Object ID} &
\colhead{R.A.} &
\colhead{Decl.} &
\colhead{$M_{\mathrm{UV}}$} &
\colhead{$z_{\mathrm{phot}}$} &
\colhead{$z_{\mathrm{spec}}$} &
\colhead{Spectroscopic Program}
}

\startdata
0303+0060 & 11997 & 45.719212 & -0.956818 & $-20.98^{+0.12}_{-0.10}$ & $8.10^{+0.07}_{-0.55}$ & $-$ & $-$ \\
0332-2745 & 17578 & 53.037659 & -27.711735 & $-19.77^{+0.19}_{-0.12}$ & $8.17^{+-0.01}_{-1.04}$ & $-$ & $-$ \\
1526+3560 & 9252 & 231.548172 & 35.993847 & $-20.50^{+0.22}_{-0.16}$ & $8.17^{+0.18}_{-0.96}$ & $-$ & $-$ \\
0227-5319 & 1200 & 36.852478 & -53.355553 & $-20.93^{+0.11}_{-0.08}$ & $8.18^{+-0.01}_{-0.97}$ & $-$ & $-$ \\
0332-2745 & 17579 & 53.037807 & -27.711599 & $-20.43^{+0.15}_{-0.10}$ & $8.22^{+0.13}_{-1.25}$ & $-$ & $-$ \\
0332-2745 & 8569 & 53.053482 & -27.765987 & $-18.86^{+0.20}_{-0.15}$ & $8.22^{+0.13}_{-0.76}$ & $-$ & $-$ \\
0332-2745 & 9828 & 53.012547 & -27.731163 & $-20.23^{+0.04}_{-0.06}$ & $8.22^{+0.04}_{-0.24}$ & $-$ & $-$ \\
1010+2701 & 19789 & 152.471619 & 27.051905 & $-19.21^{+0.29}_{-0.23}$ & $8.22^{+0.32}_{-1.01}$ & $-$ & $-$ \\
2304-6250 & 31801 & 345.996063 & -62.816113 & $-20.48^{+0.08}_{-0.08}$ & $8.23^{+0.21}_{-0.34}$ & $-$ & $-$ \\
1010+2701 & 2444 & 152.446686 & 26.987419 & $-19.97^{+0.36}_{-0.27}$ & $8.24^{+0.30}_{-1.11}$ & $-$ & $-$ \\
0217-0508 & 1427 & 34.307549 & -5.168149 & $-19.20^{+0.17}_{-0.14}$ & $8.24^{+0.10}_{-0.95}$ & $-$ & $-$ \\
0332-2745 & 8259 & 53.050213 & -27.764677 & $-20.10^{+0.18}_{-0.16}$ & $8.25^{+0.19}_{-0.71}$ & $-$ & $-$ \\
0217-0508 & 9685 & 34.272381 & -5.125317 & $-18.26^{+0.23}_{-0.18}$ & $8.29^{+0.25}_{-0.92}$ & $-$ & $-$ \\
0217-0508 & 3135 & 34.317574 & -5.149925 & $-18.68^{+0.23}_{-0.15}$ & $8.33^{+0.21}_{-1.28}$ & $-$ & $-$ \\
0217-0509 & 6946 & 34.211567 & -5.140682 & $-19.05^{+0.21}_{-0.16}$ & $8.35^{+0.28}_{-0.72}$ & $-$ & $-$ \\
0303+0060 & 9954 & 45.741615 & -0.976950 & $-19.42^{+0.21}_{-0.21}$ & $8.36^{+0.18}_{-1.15}$ & $-$ & $-$ \\
0217-0504 & 3515 & 34.270214 & -5.040026 & $-19.50^{+0.17}_{-0.12}$ & $8.39^{+0.06}_{-1.18}$ & $-$ & $-$ \\
0217-0504 & 24295 & 34.306004 & -5.045831 & $-18.30^{+0.29}_{-0.23}$ & $8.39^{+0.44}_{-0.93}$ & $-$ & $-$ \\
1420+5253 & 2835 & 214.876129 & 52.880840 & $-19.53^{+0.12}_{-0.08}$ & $8.40^{+-0.05}_{-0.14}$ & $8.361$ & EGS, GO-4287 (PI: C. Mason) [1][2] \\
0217-0509 & 13335 & 34.218163 & -5.103996 & $-18.92^{+0.25}_{-0.16}$ & $8.42^{+0.12}_{-1.13}$ & $-$ & $-$ \\
0217-0508 & 9763 & 34.272213 & -5.124932 & $-20.62^{+0.03}_{-0.03}$ & $8.52^{+0.12}_{-0.26}$ & $7.995$ & UDS, MoM (GO-5224, PI: P. Oesch) [3] \\
0860+3857 & 14145 & 134.982574 & 38.985649 & $-20.91^{+0.17}_{-0.12}$ & $8.54^{+0.09}_{-1.25}$ & $-$ & $-$ \\
0217-0508 & 8415 & 34.292923 & -5.120759 & $-18.87^{+0.32}_{-0.12}$ & $8.56^{+0.17}_{-0.76}$ & $-$ & $-$ \\
0217-0508 & 8040 & 34.302929 & -5.117467 & $-19.20^{+0.12}_{-0.11}$ & $8.58^{+0.15}_{-0.51}$ & $-$ & $-$ \\
2325-1203 & 492 & 351.319855 & -12.080091 & $-18.72^{+0.35}_{-0.59}$ & $8.62^{+1.58}_{-0.63}$ & $-$ & $-$ \\
0014-3025 & 10503 & 3.624762 & -30.437508 & $-20.71^{+0.18}_{-0.14}$ & $8.64^{+0.09}_{-0.29}$ & $-$ & $-$ \\
1010+2701 & 420 & 152.422150 & 26.986328 & $-21.25^{+0.12}_{-0.09}$ & $8.64^{+0.09}_{-0.48}$ & $-$ & $-$ \\
2325-1203 & 4929 & 351.310089 & -12.049022 & $-19.89^{+1.90}_{-0.17}$ & $8.65^{+0.18}_{-6.51}$ & $-$ & $-$ \\
1420+5253 & 6911 & 214.935486 & 52.899807 & $-20.12^{+0.09}_{-0.09}$ & $8.70^{+0.03}_{-0.16}$ & $-$ & $-$ \\
0217-0509 & 4602 & 34.220940 & -5.160698 & $-19.91^{+0.07}_{-0.06}$ & $8.71^{+0.03}_{-0.36}$ & $-$ & $-$ \\
1114+3235 & 4559 & 168.527191 & 32.574013 & $-20.81^{+0.16}_{-0.17}$ & $8.77^{+0.16}_{-0.51}$ & $-$ & $-$ \\
0217-0509 & 11648 & 34.212341 & -5.122705 & $-19.51^{+0.10}_{-0.09}$ & $8.81^{+0.12}_{-0.36}$ & $8.247$ & UDS, RUBIES (GO-4233, PI: A. de Graaff) [4] \\
1420+5252 & 2768 & 215.008652 & 52.868328 & $-20.96^{+0.06}_{-0.07}$ & $8.84^{+0.09}_{-0.20}$ & $8.779$ & EGS, CAPERS (GO-6368, PI: M. Dickinson) [5][6] \\
0338-0454 & 702 & 54.452179 & -4.939065 & $-19.62^{+0.41}_{-0.48}$ & $8.85^{+1.57}_{-0.96}$ & $-$ & $-$ \\
0227-5319 & 16860 & 36.850876 & -53.275326 & $-21.12^{+0.06}_{-0.07}$ & $8.86^{+0.06}_{-0.23}$ & $-$ & $-$ \\
\enddata
\tablenotetext{}{[1] \citet{whitlerDeepJWSTSpectroscopy2026}; [2] \citet{chenImpactGalaxyOverdensities2026}; [3] \citet{naiduCosmicMiracleRemarkably2026}; [4] \citet{degraaffRUBIESCompleteCensus2025}; [5] \citet{kokorevCAPERSObservationsTwo2025}; [6] \citet{donnanVeryBrightVery2025}}
\label{tab:f090w_dropouts_4}
\end{deluxetable*}


\begin{deluxetable*}{llcccccc}
\renewcommand{\arraystretch}{1.5} 
\caption{F090W-dropouts (continued)}

\tablehead{
\colhead{Field ID} &
\colhead{Object ID} &
\colhead{R.A.} &
\colhead{Decl.} &
\colhead{$M_{\mathrm{UV}}$} &
\colhead{$z_{\mathrm{phot}}$} &
\colhead{$z_{\mathrm{spec}}$} &
\colhead{Spectroscopic Program}
}

\startdata
0217-0504 & 10016 & 34.283627 & -5.044103 & $-19.31^{+0.21}_{-0.11}$ & $8.87^{+0.06}_{-0.43}$ & $-$ & $-$ \\
1420+5253 & 6115 & 214.940674 & 52.892456 & $-19.89^{+0.08}_{-0.08}$ & $8.89^{+-0.06}_{-0.16}$ & $-$ & $-$ \\
1010+2701 & 18712 & 152.439987 & 27.060465 & $-20.38^{+0.23}_{-0.26}$ & $8.94^{+0.82}_{-0.49}$ & $-$ & $-$ \\
0217-0504 & 7157 & 34.309956 & -5.090462 & $-20.37^{+0.08}_{-0.09}$ & $8.96^{+0.07}_{-0.23}$ & $-$ & $-$ \\
0338-0454 & 10032 & 54.466843 & -4.873794 & $-20.80^{+0.10}_{-0.10}$ & $9.00^{+0.12}_{-0.27}$ & $-$ & $-$ \\
0217-0509 & 6998 & 34.223602 & -5.134971 & $-18.43^{+0.29}_{-0.23}$ & $9.07^{+0.58}_{-0.91}$ & $-$ & $-$ \\
0217-0508 & 3296 & 34.305820 & -5.154357 & $-20.02^{+0.05}_{-0.04}$ & $9.16^{+0.17}_{-0.23}$ & $9.248$ & UDS, RUBIES (GO-4233, PI: A. de Graaff) [1] \\
0217-0508 & 7634 & 34.305424 & -5.118019 & $-18.87^{+0.25}_{-0.19}$ & $9.18^{+0.79}_{-0.54}$ & $-$ & $-$ \\
1010+2701 & 17114 & 152.439484 & 27.056280 & $-20.91^{+0.09}_{-0.10}$ & $9.25^{+0.39}_{-0.23}$ & $-$ & $-$ \\
0014-3025 & 13872 & 3.617229 & -30.425524 & $-22.16^{+0.04}_{-0.03}$ & $9.28^{+-0.05}_{-0.16}$ & $9.319$ & Abell-2744, UNCOVER (GO-2561, PI: I. Labbé) [2][3][4][5] \\
\enddata
\tablenotetext{}{[1] \citet{degraaffRUBIESCompleteCensus2025}; [2] \citet{bezansonJWSTUNCOVERTreasury2024}; [3] \citet{priceUNCOVERSurveyFirst2025}; [4] \citet{fujimotoUNCOVERNIRSpecCensus2024}; [5] \citet{boyettMassiveInteractingGalaxy2024}}
\label{tab:f090w_dropouts_5}
\end{deluxetable*}

\begin{deluxetable*}{llcccccc}
\renewcommand{\arraystretch}{1.5} 
\caption{F115W-dropouts}

\tablehead{
\colhead{Field ID} &
\colhead{Object ID} &
\colhead{R.A.} &
\colhead{Decl.} &
\colhead{$M_{\mathrm{UV}}$} &
\colhead{$z_{\mathrm{phot}}$} &
\colhead{$z_{\mathrm{spec}}$} &
\colhead{Spectroscopic Program}
}

\startdata
0860+3857 & 10356 & 134.977585 & 38.971680 & $-19.56^{+0.42}_{-0.41}$ & $8.97^{+1.22}_{-0.34}$ & $-$ & $-$ \\
0217-0509 & 11058 & 34.217091 & -5.122624 & $-18.15^{+0.42}_{-0.33}$ & $9.13^{+1.29}_{-0.20}$ & $-$ & $-$ \\
0217-0509 & 12959 & 34.221210 & -5.113707 & $-17.81^{+0.70}_{-0.65}$ & $9.44^{+0.97}_{-0.91}$ & $-$ & $-$ \\
0332-2749 & 16743 & 53.001728 & -27.812489 & $-19.58^{+0.17}_{-0.16}$ & $9.48^{+0.83}_{-0.25}$ & $-$ & $-$ \\
0332-2749 & 19513 & 53.075996 & -27.806505 & $-19.73^{+0.14}_{-0.12}$ & $9.91^{+0.51}_{-0.47}$ & $-$ & $-$ \\
1010+2701 & 12141 & 152.432724 & 27.044855 & $-18.46^{+0.44}_{-0.48}$ & $9.91^{+0.85}_{-0.79}$ & $-$ & $-$ \\
0332-2749 & 17529 & 53.085506 & -27.808601 & $-19.90^{+0.12}_{-0.11}$ & $10.11^{+0.54}_{-0.78}$ & $-$ & $-$ \\
0217-0509 & 13052 & 34.221394 & -5.113218 & $-19.71^{+0.23}_{-0.25}$ & $10.17^{+0.13}_{-0.84}$ & $-$ & $-$ \\
0227-5319 & 15603 & 36.824085 & -53.289127 & $-20.01^{+0.36}_{-0.42}$ & $10.31^{+0.11}_{-1.18}$ & $-$ & $-$ \\
2304-6250 & 25548 & 345.971710 & -62.829235 & $-20.97^{+0.15}_{-0.08}$ & $10.57^{+0.44}_{-0.60}$ & $-$ & $-$ \\
0959+0200 & 14138 & 149.845947 & 2.032689 & $-19.95^{+0.18}_{-0.17}$ & $10.72^{+0.05}_{-1.59}$ & $-$ & $-$ \\
0332-2749 & 6608 & 53.005013 & -27.824554 & $-19.37^{+0.28}_{-0.25}$ & $11.02^{+2.21}_{-1.79}$ & $-$ & $-$ \\
0015-3034 & 7618 & 3.668404 & -30.548479 & $-20.21^{+0.12}_{-0.13}$ & $11.50^{+0.25}_{-0.61}$ & $-$ & $-$ \\
1420+5253 & 105 & 214.861679 & 52.861858 & $-19.16^{+0.17}_{-0.17}$ & $11.91^{+0.36}_{-0.66}$ & $11.471$ & EGS, GO-4106 (PI: E. Nelson) [1] \\
\enddata
\tablenotetext{}{[1] \citet{rodighieroEGSz11R0RedDustrich2026}}
\label{tab:f115w_dropouts_1}
\end{deluxetable*}

\end{appendix}
\end{document}